\documentclass[conference,hyphens]{IEEEtran}

\usepackage[hyphens]{url}
\usepackage{cite}
\usepackage{hyperref}
\usepackage{wrapfig,booktabs}
\hypersetup{
    colorlinks=true,
    linkcolor=blue,
    filecolor=blue,      
    urlcolor=blue,
}
\usepackage{makecell}
\usepackage{amsmath,amssymb,amsfonts}
\usepackage{siunitx}
\usepackage{algorithmic}
\usepackage{graphicx}
\usepackage{textcomp}
\usepackage{xcolor}
\usepackage{multirow}
\usepackage{tabularx} 
\usepackage{soul}
\usepackage{orcidlink}
\usepackage{listings}
\usepackage{pythonhighlight}

\makeatletter
\renewcommand\@IEEEsectpunct{\ } 
\makeatother

\usepackage[binary-units]{siunitx}
\usepackage{tikz}
\usepackage{pgfplots}
\usepackage{scalefnt}
\pgfplotsset{compat=1.17} %
\usepgfplotslibrary{statistics} %

\pgfplotsset{compat=1.11,
    /pgfplots/ybar legend/.style={
        /pgfplots/legend image code/.code={%
            \draw[##1,/tikz/.cd,yshift=-0.25em] (0cm,0cm) rectangle (3pt,0.8em);
        },
    },
}

\definecolor{s1}{RGB}{228, 26, 28}
\definecolor{s2}{RGB}{55, 126, 184}
\definecolor{s3}{RGB}{77, 175, 74}
\definecolor{s4}{RGB}{152, 78, 163}
\definecolor{s5}{RGB}{255, 127, 0}
\definecolor{s6}{RGB}{25, 127, 0}
\definecolor{s7}{RGB}{25, 127, 55}
\definecolor{s8}{RGB}{255, 127, 45}
\definecolor{s9}{RGB}{85, 127, 65}
\definecolor{s10}{RGB}{0, 0, 255}
\definecolor{s11}{RGB}{0, 0, 0}
\definecolor{s12}{RGB}{255, 0, 0}

\pgfplotscreateplotcyclelist{set1}{
    s11,every mark/.append style={fill=s11},mark=*\\
    s2,every mark/.append style={fill=s2},mark=*\\
    s3,every mark/.append style={fill=s3},mark=triangle*\\
    s4,every mark/.append style={fill=s4},mark=square\\
    s5,every mark/.append style={fill=s5},mark=diamond*\\
}

\pgfplotscreateplotcyclelist{set111}{
    s1,every mark/.append style={fill=s1},mark=*\\
    s2,every mark/.append style={fill=s2},mark=*\\
    s3,every mark/.append style={fill=s3},mark=triangle*\\
    s4,every mark/.append style={fill=s4},mark=square\\
    s5,every mark/.append style={fill=s5},mark=diamond*\\
}

\pgfplotscreateplotcyclelist{set1111}{
    s3,every mark/.append style={fill=s3},mark=*\\
    s2,every mark/.append style={fill=s2},mark=*\\
    s3,every mark/.append style={fill=s3},mark=triangle*\\
    s4,every mark/.append style={fill=s4},mark=square\\
    s5,every mark/.append style={fill=s5},mark=diamond*\\
}

\pgfplotscreateplotcyclelist{set2}{
    s1, style={fill=s1}\\
    s2, style={fill=s2}\\
    s8, style={fill=s3}\\
    s9, style={fill=s9}\\
}

\pgfplotscreateplotcyclelist{set28}{
    s10, style={fill=s10}\\
    s8, style={fill=s8}\\
    s9, style={fill=s9}\\
}

\usetikzlibrary{patterns}

\pgfplotscreateplotcyclelist{set3}{
    {s1, fill=s1}, %
    {s2, fill=s2}, 
    {s3, fill=s3}, 
    {s4, fill=s4}, 
    {s5, fill=s5}
}

\pgfplotscreateplotcyclelist{set4}{
    s10,every mark/.append style={fill=s10},mark=*,mark size=1.0pt\\
    s3,every mark/.append style={fill=s3},mark=*,mark size=1.0pt\\
    s1,every mark/.append style={fill=s1},mark=*,mark size=1.0\\
    s2,every mark/.append style={fill=s2},mark=o\\
    s3,every mark/.append style={fill=s3},mark=triangle*\\
    s4,every mark/.append style={fill=s4},mark=square\\
    s5,every mark/.append style={fill=s5},mark=diamond*\\
}

\pgfplotscreateplotcyclelist{set18}{
    s1,every mark/.append style={fill=s1},mark=o,mark size=0.7pt\\
    s2,every mark/.append style={fill=s2},mark=square,mark size=0.7pt\\
    s3,every mark/.append style={fill=s3},mark=triangle,mark size=0.7pt\\
    s4,every mark/.append style={fill=s4},mark=o,mark size=1.3pt\\
    s5,every mark/.append style={fill=s5},mark=diamond*,mark size=1.0pt\\
}

\pgfplotscreateplotcyclelist{set14}{
    s6,every mark/.append style={fill=s6}\\
    s5,every mark/.append style={fill=s5},mark=triangle,mark size=1.3pt, only marks\\
    s2,every mark/.append style={fill=s2},mark=o,mark size=1.0pt\\
    s3,every mark/.append style={fill=s3},mark=triangle,mark size=2.0pt\\
    s4,every mark/.append style={fill=s4},mark=square,mark size=1.5pt, only marks\\
    s5,every mark/.append style={fill=s5},mark=diamond*,mark size=1.0pt\\
}

\pgfplotscreateplotcyclelist{set24}{
    s1,every mark/.append style={fill=s1},mark=*,mark size=1.0pt, only marks\\
    s2,every mark/.append style={fill=s2},mark=o,mark size=1.0pt, only marks\\
    s3,every mark/.append style={fill=s3},mark=triangle*,mark size=1.0pt, only marks\\
    s4,every mark/.append style={fill=s4},mark=square,mark size=1.5pt, only marks\\
    s5,every mark/.append style={fill=s5},mark=diamond*,mark size=1.0pt, only marks\\
}

\pgfplotscreateplotcyclelist{set28}{
    s1,every mark/.append style={fill=s1},mark=*,mark size=2.5pt, only marks\\
    s2,every mark/.append style={fill=s2},mark=o,mark size=2.5pt, only marks\\
    s3,every mark/.append style={fill=s3},mark=triangle*,mark size=2.5pt, only marks\\
    s4,every mark/.append style={fill=s4},mark=square,mark size=2.5pt, only marks\\
    s5,every mark/.append style={fill=s5},mark=diamond*,mark size=2.5pt, only marks\\
}

\usepackage[caption=false]{subfig}
\def\BibTeX{{\rm B\kern-.05em{\sc i\kern-.025em b}\kern-.08em
    T\kern-.1667em\lower.7ex\hbox{E}\kern-.125emX}}

\newcommand{\up}{\vspace*{-1em}}
\newcommand{\upp}{\vspace*{-0.5em}}

\newif\ifdraft
\ifdraft
\newcommand{\IPDPS}[1]{\textcolor{red} {**IPDPS Review: #1}} 
\newcommand{\CCGRID}[1]{\textcolor{brown} {**CCGRID Review: #1}}
\newcommand{\note}[1]{ {\textcolor{orange} { **Note: #1 }}}
\newcommand{\jhanote}[1]{ {\textcolor{red} { ***Shantenu: #1 }}}
\newcommand{\alnote}[1]{ {\textcolor{orange} { ***Andre: #1 }}}
\newcommand{\nsnote}[1]{ {\textcolor{green} { ***Nishant: #1 }}} 
\newcommand{\pnote}[1]{ {\textcolor{green} { ***Pradeep: #1 }}}
\newcommand{\fknote}[1]{ {\textcolor{blue} { ***Florian: #1 }}}
\else
\newcommand{\note}[1]{}
\newcommand{\alnote}[1]{}
\newcommand{\nsnote}[1]{}
\newcommand{\jhanote}[1]{}
\newcommand{\crnote}[1]{}
\newcommand{\jknote}[1]{}
\newcommand{\pnote}[1]{}
\newcommand{\fknote}[1]{}
\newcommand{\IPDPS}[1]{}
\newcommand{\CCGRID}[1]{}
\fi
\newcommand{\ignore}[1]{}

\lstdefinestyle{custompython}{
    backgroundcolor=\color{white},
    basicstyle=\footnotesize\ttfamily,
    breaklines=true,
    frame=single,
    numbers=left,
    numberstyle=\tiny\color{gray},
    keywordstyle=\color{blue},
    commentstyle=\color{green},
    stringstyle=\color{red},
}

\begin{document}
\bstctlcite{IEEEexample:BSTcontrol}
\title{Pilot-Quantum: A Middleware for Quantum-HPC Resource, Workload and Task Management}

\author{\IEEEauthorblockN{Pradeep Mantha\orcidlink{0000-0003-2664-7737}\IEEEauthorrefmark{1}, Florian J. Kiwit\orcidlink{0009-0000-4065-1535}\IEEEauthorrefmark{1}\IEEEauthorrefmark{6}, Nishant Saurabh\orcidlink{0000-0002-1926-4693}\IEEEauthorrefmark{2}, Shantenu Jha\orcidlink{0000-0002-5040-026X}\IEEEauthorrefmark{3}\IEEEauthorrefmark{4}\IEEEauthorrefmark{5}, Andre Luckow\orcidlink{0000-0002-1225-4062}\IEEEauthorrefmark{1}\IEEEauthorrefmark{6}}
\IEEEauthorblockA{\IEEEauthorrefmark{1}{\textit{Ludwig Maximilian University, Munich, Germany}}\\
\IEEEauthorrefmark{2}\textit{Department of Information and Computing Sciences, Utrecht University, NL}\\
\IEEEauthorrefmark{3}{\textit{Rutgers University-New Brunswick, New Jersey, USA}}\\
\IEEEauthorrefmark{4}{\textit{Princeton Plasma Physics Laboratory, New Jersey, USA}}\\
\IEEEauthorrefmark{5}{\textit{Princeton University, New Jersey, USA}}\\
\IEEEauthorrefmark{6}{\textit{BMW Group, Munich, Germany}}\\
pradeepm66@gmail.com\IEEEauthorrefmark{1},
f.kiwit@campus.lmu.de\IEEEauthorrefmark{1},
n.saurabh@uu.nl\IEEEauthorrefmark{2},
shantenu.jha@rutgers.edu\IEEEauthorrefmark{3},\\
a.luckow@lmu.de\IEEEauthorrefmark{1}
}\upp\upp\upp\upp\upp
}


\maketitle

\begin{abstract}

As quantum hardware advances, integrating quantum processing units (QPUs) into HPC environments and managing diverse infrastructure and software stacks becomes increasingly essential. \textit{Pilot-Quantum} addresses these challenges as a middleware designed to provide unified application-level management of resources and workloads across hybrid quantum-classical environments. It is built on a rigorous analysis of existing quantum middleware systems and application execution patterns. It implements the Pilot Abstraction conceptual model, originally developed for HPC, to manage resources, workloads, and tasks. It is designed for quantum applications that rely on task parallelism, including (i) \textit{Hybrid algorithms}, such as variational approaches, and (ii) \textit{Circuit cutting systems}, used to partition and execute large quantum circuits. \textit{Pilot-Quantum} facilitates seamless integration of QPUs, classical CPUs, and GPUs, while supporting high-level programming frameworks like Qiskit and Pennylane. This enables users to efficiently design and execute hybrid workflows across diverse computing resources. The capabilities of \textit{Pilot-Quantum} are demonstrated through mini-apps -- simplified yet representative kernels focusing on critical performance bottlenecks. We demonstrate the capabilities of Pilot-Quantum through multiple mini-apps, including different circuit execution (e.g., using IBM’s Eagle QPU and simulators), circuit-cutting, and quantum machine learning scenarios.
\end{abstract}


\section{Introduction}


\alnote{Objective: enable application/workload-aware scheduling on hybrid quantum/hpc infrastructures}

\alnote{speak to the challenges of increasingly heterogeneous hpc computing as classical, accelerated/specialized, and quantum computing converge}

\alnote{reference Suttons bitter lesson as motivation for Pilot quantum as enabler and middleware for bringing large scale, heterogeneous compute to learning algorithms and applications.}

\alnote{speak about the different types of parallelism: wide circuit with max. number of qubits vs. max parallelism with heterogeneous smaller circuits}

\alnote{challenges for quantum-classical hybrid workflows: mastering multiple programming paradigms - solution abstraction, optimization algorithms for different QPUs and simulation environments (CPU, GPU)}

\alnote{speak to philosophy of handling resource and scheduling decision at lowest possible layer}


\alnote{why pilot? what is the difference between pilot and qiskit? Concept, example, experiment? 2 Arguments: (i) Pilot is an overlay on Qiskit for going distributed (distinguish resource vs. runtime overlay and (ii) application-level scheduling}\jhanote{I think these are fundamental and critical points to address} \alnote{should be addressed in the second to last paragraph of the introduction section now}

\textit{Motivation:} Quantum computing promises to advance specific science and industry computational challenges, from chemistry, optimization and simulations to machine learning~\cite{bayerstadler2021industry}, by theoretically outperforming classical methods by requiring fewer computational steps~\cite{nielsen_chuang_2010,dalzell2023quantum}. While quantum hardware has progressed, e.\,g., concerning qubit fidelity~\cite{quantinuum_h2} and error correction~\cite{acharya2024quantumerrorcorrectionsurface}, further improvements are needed for a practical quantum advantage~\cite{quantumcomputingautomotiveapplications}.

\textit{Quantum-HPC Integration:} To realize the benefits of quantum computing, tighter integration of classical and quantum resources will become increasingly important~\cite{ALEXEEV2024666, BECK202411, mohseni2024buildquantumsupercomputerscaling, RUEFENACHT2022, hpcqc}. Such integration presents challenges across several levels: (i) \textit{Hardware Integration}: Efficiently combining heterogeneous quantum systems (e.g., superconducting and ion-trap qubits) with classical computing resources (e.\,g., GPUs and CPUs) to create cohesive hybrid systems.
(ii) \textit{Software Ecosystems}: Developing robust software frameworks, tools, and middleware systems that seamlessly integrate with existing HPC tools (e.\,g., CUDA and MPI runtimes). Achieving this integration is crucial for advancing quantum computing and unlocking its full potential in solving complex computational problems.

\textit{Challenges:} Resource, workload, and task management in classical workflow systems face significant challenges arising from heterogeneity, application and infrastructure dynamism, complex integration needs, and software fragmentation~\cite{alsaadi2024exascaleworkflowapplicationsmiddleware}. While quantum-HPC systems share these difficulties, they substantially amplify them due to additional complexities unique to quantum computing. For example, leveraging quantum-HPC systems fully requires integrating multiple new types of infrastructure and support for new parallelism and integration patterns, e.g., circuit-level parallelism across multiple QPUs, tightly-coupled GPU-QPU integration for error correction, and loosely-coupled integration of variational quantum algorithms (VQA). Additional new task constraints, including qubit counts, fidelity thresholds, and classical-quantum resource co-location, must also be considered for scheduling and execution. Further, the rapidly evolving and often fragmented quantum hardware, software libraries \cite{elsharkawy2023integration, javadiabhari2024quantumcomputingqiskit,bergholm2022pennylaneautomaticdifferentiationhybrid}, and middleware \cite{quantum_serverless_qsw,covalent_zenodo,CUDAQ} landscape makes integration and scaling difficult.


\emph{Quantum-HPC Middleware:} Modern workflow systems are increasingly shifting from monolithic designs to modular architectures that separate concerns across distinct layers~\cite{alsaadi2024exascaleworkflowapplicationsmiddleware}. Following this principle, we introduce Pilot-Quantum, a modular middleware for resource, workload, and task management that provides an application-level framework for orchestrating hybrid quantum-classical workloads across CPUs, GPUs, and QPUs. Pilot-Quantum addresses the key challenges of hybrid execution through multi-level scheduling, enabling adaptive resource allocation tailored to application needs and efficient use of costly and heterogeneous resources. Pilot-Quantum builds upon the Pilot-Abstraction~\cite{6404423,10.1145/3177851}, a conceptual model for quantum-HPC middleware~\cite{saurabh2023conceptual} and an analysis of execution motifs~\cite{saurabh2024quantum}. It supports integration with high-level frameworks like Qiskit~\cite{javadiabhari2024quantumcomputingqiskit,Qiskit_OpenSource} and PennyLane~\cite{bergholm2022pennylaneautomaticdifferentiationhybrid} and low-level runtimes like cuQuantum~\cite{bayraktar2023cuquantum}. This design enables scalable, efficient execution of complex hybrid workflows in rapidly evolving quantum-HPC environments.

This paper is structured as follows: We explore quantum software and middleware state-of-the-art in section~\ref{sec:related}. The primary contribution is Pilot-Quantum, an open-source middleware~\cite{Mantha_Pilot-Quantum_2024} addressing resource, workload, and task management challenges, introduced in section~\ref{sec:middleware}. Section~\ref{sec:eval} demonstrates its capabilities by extending quantum mini apps~\cite{saurabh2024quantum} introducing, e.g., a circuit cutting and quantum machine learning (QML) mini-apps. We conclude with a qualitative comparison of quantum middleware systems with Pilot-Quantum and future work in section~\ref{sec:discussion_conclusion}.

\section{Background and Related Work}
\label{sec:related}
This section analyzes quantum-classical software and middleware using a conceptual architecture model~\cite{saurabh2023conceptual}, quantum-HPC integration modes and execution motifs~\cite{saurabh2024quantum}.
\subsection{Conceptual Middleware Architecture}
\label{sec:conceptual}
In~\cite{saurabh2023conceptual}, we define a conceptual architecture for a high-level system design of quantum-HPC middleware systems that explains the main components and their interactions. It consists of four distinct layers, denoted as $L1$, $L2$, $L3$, $L4$. $L4$ is the workflow layer, which encapsulates application semantics and logical dependencies between the workflow stages and offers a high-level abstraction for developing the quantum-HPC workflows. The workload layer $L3$ manages the workload, a set of interdependent, quantum, and classical tasks emitted by $L4$. It optimizes the workload execution by partitioning tasks and assigning them to the available resources utilizing parallelism (e.g., ensemble and task parallelism).

The $L2$ task layer performs the operational management of task execution on a single resource. Any runtime operations, i.\,e., task scheduling, assignment, execution, and monitoring, are managed at $L2$. Finally, the resource layer $L1$ describes how a system abstracts and manages heterogeneous classical (e.\,g., HPC and cloud systems, including accelerators) and quantum resources (e.\,g., gate-based and analog QPUs). $L1$ layer provides the capability to exploit system-level tools, e.\,g., GPU acceleration and QPU-target optimization are essential for efficient quantum-classical integration.

\subsection{Quantum-HPC Integration Modes and Motifs}

\IPDPS{Define task and workload here?}

Understanding common quantum-HPC integration modes and motifs (i.e., recurring execution patterns) between quantum-HPC components~\cite{saurabh2024quantum} is critical to designing effective middleware systems. 

\paragraph*{Integration modes} are categorized as \emph{Quantum-x-HPC} or \emph{HPC-x-Quantum} based on the strength of coupling between quantum and classical tasks~\cite{ALEXEEV2024666, saurabh2023conceptual}. Three modes are identified: \emph{HPC-for-Quantum}, with strong coupling optimizing quantum-classical interactions (e.\,g., error mitigation~\cite{ella2023quantumclassical}, error correction~\cite{hong2024entanglinglogicalqubitsbreakeven, Bluvstein_2023}, dynamic circuits~\cite{PhysRevLett.127.100501}); \emph{Quantum-in-HPC} emphasizes loosely-coupled parallelism between classical and quantum resources found in VQA~\cite{vqa}  to accelerate HPC applications; and \emph{Quantum-about-HPC}, with loose coupling of quantum elements into HPC applications for pre- and post-processing~\cite{generative_molecule_design_2022, https://doi.org/10.48550/arxiv.2101.06250}.


\paragraph*{Motifs}~\cite{saurabh2024quantum} are recurring high-level execution patterns describing quantum-classical task interactions, e.\,g., whether operations occur within the quantum system's coherence time (for tightly coupled tasks) or beyond it (for loosely coupled tasks). For classical simulation of quantum systems, both coupling components exist. Distributed state vector simulation involves tightly coupled classical tasks using MPI and GPU/CUDA parallelism, while variational algorithms and circuit cutting typically require loose coupling between tasks.

\subsection{Quantum Software Layers}
\label{sec_qsw}

\begin{figure}[t]
    \centering
    \includegraphics[width=0.8\columnwidth]{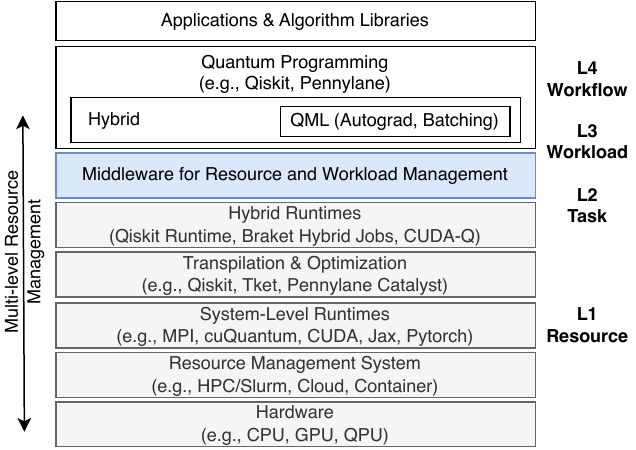}\upp\upp
    \caption{\textbf{Quantum Software Stack:} An overview of key components and layers, from quantum programming environments and hybrid runtimes to hardware resource management.}
    \label{fig:quantum-software}
    \upp \upp
\end{figure}

This section details the quantum software stack, mapping it to the layers of the conceptual model (see Figure~\ref{fig:quantum-software} and~\cite{elsharkawy2023integration}).



\paragraph*{Quantum Programming} frameworks typically provide quantum circuit model-based abstraction for describing quantum programs. Examples are Qiskit~\cite{javadiabhari2024quantumcomputingqiskit,Qiskit_OpenSource}, PennyLane~\cite{bergholm2022pennylaneautomaticdifferentiationhybrid}, Cirq~\cite{Cirq2024},  low-level frameworks like CUDA-Q and AWS Braket~\cite{braket}, and high-level programming libraries, such as Qrisp~\cite{seidel2024qrispframeworkcompilablehighlevel}, Silq~\cite{10.1145/3385412.3386007}, and Q\#~\cite{Svore_2018}. At the $L4+$ levels, domain-specific programming libraries include Qiskit Optimization~\cite{qiskit_optimization}, Qiskit Nature~\cite{qiskit_nature}, and OpenFermion~\cite{mcclean2019openfermionelectronicstructurepackage}.


\paragraph*{Hybrid Runtimes} like Qiskit Runtime~\cite{Qiskit_IBM_Runtime} and Braket Hybrid Jobs~\cite{AWSBraketJobs,braket-jobs-2021} provide session and decorator abstractions, respectively, to co-allocate heterogeneous resources for hybrid applications. However, new execution patterns, such as the combination of quantum and machine learning capabilities, require new abstractions, including auto differentiation of quantum circuits (e.\,g., using PennyLane~\cite{bergholm2022pennylaneautomaticdifferentiationhybrid}) or training classical models using quantum task outputs~\cite{nakaji2024generative}. Some frameworks like XACC~\cite{xacc_2020}, QCOR~\cite{nguyen2020extendingcheterogeneousquantumclassical}, and CUDA-Q~\cite{CUDAQ} use a kernel-based programming model with compile toolchains to manage quantum-classical function interactions. For classical simulation of quantum systems, acceleration frameworks, e.\,g., cuQuantum~\cite{bayraktar2023cuquantum} are critical for performance and scale.

\paragraph*{Hardware Integration, Transpiler/Compiler, and Optimization:} 
At the $L1$ layer, many quantum frameworks provide device abstractions and plugins for accessing quantum devices, such as Qiskit backends~\cite{QiskitBackends2023}, Qiskit Runtime~\cite{Qiskit_IBM_Runtime}, and PennyLane plugins~\cite{pennylane_plugins}. Intermediate representations like OpenQASM~\cite{Cross_2022} and QIR~\cite{QIRAlliance, lubinski2022advancinghybridquantumclassicalcomputation} aim to bridge the gap between hardware platforms and high-level programming constructs. Transpilers and compilers are used to transform a circuit into a hardware-optimized form. They are provided by full-stack frameworks like Qiskit and specialized frameworks like Tket~\cite{tket}, BQSKit~\cite{osti_1785933}, and Staq~\cite{staq} transpilers.


\paragraph*{Multi-level Resource Management:}
Resource management and scheduling in quantum-HPC environments occur across multiple levels and time scales~\cite{BECK202411, elsharkawy2024integrationquantumacceleratorshpc}, involving integration of coarse-grained job-level allocation (e.\,g., GPUs) with QPU allocations and fine-grained, application-level scheduling~\cite{10.1145/3177851, AHN2020202}. Existing HPC resource management frameworks often assume exclusive resource access, leading to resource fragmentation and underutilization~\cite{288717}. 
As algorithms scale across heterogeneous resources (e.\,g., CPUs, GPUs, QPUs), such resource management will result in suboptimal performance and utilization. Pilot-Quantum specifically addresses the need for an application-level resource management by balancing system and application objectives.

\paragraph*{Middleware and System-level Runtimes} at the task and workload layer ($L3$ and $L4$) are needed to orchestrate quantum workloads across integrated quantum-HPC systems. QML workloads, for instance, involve heterogeneous classical-quantum tasks for gradient computation, backpropagation, optimization, and circuit execution, requiring a systematic approach to workload and task management.

\subsection{Middleware for Resource, Workload and Task Management}
\label{sec:related_middleware}

Many quantum middleware systems emerged recently including, QFaaS~\cite{nguyen2024qfaas}, Tierkreis~\cite{https://doi.org/10.48550/arxiv.2211.02350}, Qibo~\cite{qibo_paper}, Quantum Framework (QFw)~\cite{shehata2024frameworkintegratingquantumsimulation}, Quantum Brilliance~\cite{Nguyen2024}, Rasqal~\cite{rasqal} and Zapata's Orquestra~\cite{zapata2021orchestra}. We particularly investigate Qiskit Serverless~\cite{quantum_serverless_qsw}, Covalent~\cite{covalent_zenodo}, and CUDA-Q~\cite{CUDAQ} systems. 

\subsubsection{Qiskit Serverless~\cite{ibm_quantum_serverless_2023, IBMQiskitServerless, quantum_serverless_qsw, quantum_serverless_0.17.0}}\label{sec:qiskit_serverless} is a middleware for managing resources and workloads in hybrid quantum-classical infrastructures while supporting long-running applications and scaling across distributed resources (CPU, GPU, QPU). It extends cloud-based quantum hardware access to classical computing environments at the $L2$ and $L3$ layers, utilizing Apache Ray~\cite{ray} to execute hybrid workloads. 

Qiskit Serverless maps to the $L2$/$L3$ layer of the conceptual architecture, facilitating loosely and tightly coupled interactions. For workload management at the $L3$ layers, it offers a \textit{distributed task} decorator for manual task-resource mapping. It interfaces with IBM Quantum Cloud services and hardware providers like IonQ and Braket~\cite{qiskit_ionq_provider,qiskit_braket_provider} through Qiskit Runtime and backend APIs. QPU priority is available for computations within the same Qiskit Runtime $Session$, allowing for the co-location of quantum and classical tasks.

\label{sec:covalent} 
\subsubsection{Covalent~\cite{covalent_zenodo, covalent2023}} \label{sec:covalent} 
orchestrates distributed computing workflows across Cloud-HPC resources. Its SDK provides $L4$ level abstraction using \emph{electron} and \emph{lattice} decorators for defining tasks and workflows (i.\,e., DAG of \emph{electrons}), where quantum tasks are defined using \emph{qelectron} decorator. The \emph{Dispatch} service at the $L3$ layer accepts workloads (i.\,e., ready-to-run tasks), maps them to suitable resources, and dispatches them to appropriate backends.

At the $L2$ and $L1$ layers, Covalent uses \textit{executors} to abstract resources, e.\,g., so-called \texttt{QExecutors} for quantum resources. Covalent allows grouping executors in structures like \textit{QCluster} (i.\,e., set of quantum resources), enabling parallelism within resources managed by one executor (e.g., GPU acceleration) and across multiple executors. Covalent uses Dask~\cite{dask} as the default task manager and necessitates manual resource configuration, including AWS Dask cluster setup via Terraform, which lacks HPC infrastructure support. Limited HPC support, e.\,g., via a pilot plugin~\cite{covalent_hpc_plugin} for Radical-Pilot/PSI/J, exists. For application-level task-to-executor mapping, Covalent's dispatch service considers \emph{electron} decorator-specified requirements and executor availability for dynamic task assignment at runtime. However, limited quantum cloud provider support by Covalent prevents co-scheduling of classical and quantum electrons, hindering QPU prioritization.


\subsubsection{CUDA-Q}%
~\cite{CUDAQ} supports tight integrations within heterogeneous computing environments. It employs an LLVM compiler toolchain~\cite{llvm} to convert frontend code in C++ or Python to different heterogeneous backends, such as CPUs, CUDA-enabled GPUs, and QPUs. It utilizes distinct intermediate representations for each processor type (e.\,g., QIR for QPUs and NVVM IR for GPUs), allowing for targeted optimizations. The CUDA-Q runtime, utilizes MPI for parallel processing and CUDA as a low-level runtime. While particularly suitable for \emph{HPC-for-Quantum} tight integration, CUDA-Q also supports \emph{Quantum-for-HPC} modes, e.\,g., for VQAs. At $L4$, it provides a circuit-based programming model and a kernel API for offloading circuits to hardware or simulated QPUs. Application-level task management uses an MPI and CUDA-Runtime on the $L2$/$L1$ levels.

\section{Pilot-Quantum Design and Architecture}
\label{sec:middleware}

\IPDPS{Several important things for the quantum-HPC middleware seem to be omitted in the paper. For example, if an application uses cloud services, how should user authentication be implemented for the services?}

\IPDPS{For task-based application programming, the way the program is described and the dependency between tasks are important. However, the role of the Pilot and that of a user in task-based app programming are not clear.}


\alnote{emphasize: pilot-abstraction as runtime system for heterogeneous tasks. } 
\alnote{emphasize the need for a task/resource management abstraction that is framework independent}
\alnote{describe the integration of pilot-quantum with workflow level (e.g., other high-level tools) and system level (e.\,g., different simulators)}


\IPDPS{Could you define the terms "workload" and "task" more precisely in an early section of the paper? It is described in P.7, but it may be too late.}\alnote{either here or in sec II}

\emph{Pilot-Quantum} manages resources, workloads, and tasks across heterogeneous classical and quantum infrastructures for quantum applications and workflows. It is an implementation of the Pilot Abstraction~\cite{6404423,10.1145/3177851},  a conceptual abstraction for application-level scheduling~\cite{6404423, 10.1145/369028.369109}. It uses placeholder jobs for application-level resource control, enabling flexible and adaptive scaling of resources and workloads.

\emph{Abstraction for Resource Management:} Pilot-Quantum enables applications to allocate resources (i.e., nodes, cores, GPUs, and QPUs) and execute task-based workloads on them. It provides application-level control over resources, enabling adaptive optimization of resource allocations to workloads and vice versa. This allows the application to dynamically acquire or release resources at runtime, such as optimizing the mix of classical and quantum resources for a specific workflow step.
It supports typical workloads in quantum workflows, such as running circuit ensembles, decomposing larger circuits via circuit cutting, and handling data-parallel QML workloads.

\emph{Easy Deployment:} Pilot-Quantum simplifies workflow deployment across distributed, heterogeneous infrastructures, including non-local quantum clouds, HPC MPI/CUDA environments for distributed state vector simulation, and loosely coupled workflow systems. It can also integrate with resource-specific optimization tools like QPU transpilers.

\emph{Application Performance:} Pilot-Quantum uses hierarchical, multi-level resource management to align system- and application-level requirements, enabling application-level optimization of workload and resource allocation and use. This is crucial for distributed environments with application demands and varying resource availabilities. For instance, in congested QPU hardware, its multi-level scheduling balances system utilization with application workload execution requirements.

\subsection{Requirements and Design Objective}
\label{sec:design}
\IPDPS{Design Challenges: In Section III-B (Design), it would help if the authors provided more context on the challenges they encountered with their design. Additionally, elaborating on how their design stands out compared to SOTA would clarify the technical innovations and difficulties that make this work unique.}
\emph{Infrastructure Heterogeneity:} The system must address software and hardware heterogeneity, encompassing a range of QPUs like superconducting qubits, ion traps, neutral atoms, and simulated QPUs using MPI or CUDA. This requires compatibility with diverse frameworks and tools, such as transpilers, compilers, and runtimes, including simulators (e.\,g.,  state vector and tensor network simulations with MPI or CUDA) and machine learning runtimes like PyTorch and JAX.

\emph{Applications:} Quantum-HPC applications exhibit diverse workloads~\cite{saurabh2024quantum, QuantumMiniApp}, as they evolve from monolithic applications to complex multi-stage workflows~\cite{9309042}. It can involve quantum data encoding (e.\,g., amplitude encoding, Jordan-Wigner transformation for molecules), quantum mechanical system simulation (e.\,g., ground state energy), and machine learning-based error mitigation strategies. Such workflows require diverse infrastructures and resources across execution stages, whereas classical computation may require accelerators like GPUs for simulation and machine learning. 

\emph{Resource management} in quantum-classical computing is challenging due to diverse resource types and application demands. Dynamic allocation during workflow execution is crucial to address application and infrastructure dynamism when handling tasks and dependencies across heterogeneous platforms. Multi-level scheduling enables aligning application and system objectives, efficient resource allocation, workload partitioning, and task management strategies to optimize application performance and system utilization.

\emph{Abstractions:} Integrating quantum computing into HPC workflows requires flexible abstractions that hide quantum complexity while supporting various parallelism forms (e.g., task parallelism for circuits, distributed state vector simulations using MPI and cuQuantum, and differentiable computing for QML). These should facilitate efficient pipeline creation for end-to-end quantum applications with pre- and post-processing integrations. The abstractions must offer seamless interoperability with quantum and HPC infrastructures, ensuring compatibility with frameworks like CUDA for GPU-accelerated tasks and MPI for distributed computing.

\emph{Interoperability:} Rapidly evolving quantum programming frameworks across all technology stack layers necessitate low-overhead integration with existing ones at both application and system levels. For instance, it's crucial to integrate with vendor-specific low-level runtimes that enable tight coupling between quantum and classical resources, such as CUDA-Q's MPI/CUDA runtime or Qiskit Runtime.

\subsection{System Architecture}

\begin{figure}[t]
  \centering
  \includegraphics[width=0.5\textwidth]{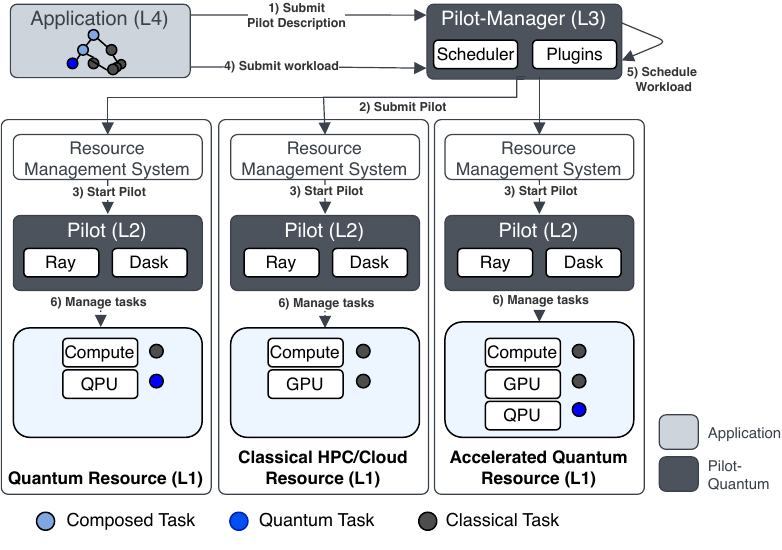}\upp \upp
  \caption{\textbf{Pilot Quantum Architecture:} The system core is a \emph{Pilot-Manager} that orchestrates and manages resources through pilots across both classical and quantum infrastructures, such as QPUs, GPUs, and CPUs. Pilots are responsible for reserving resources and managing task execution. }
  \label{fig:quantum-pilot-arch}
  \upp \upp
\end{figure}

Figure~\ref{fig:quantum-pilot-arch} shows Pilot-Quantum's architecture. The Pilot-Job system~\cite{6404423,10.1145/3177851} acts as an intermediary between applications, higher-level systems (e.g., workflow management), and local resource management systems (e.\,g., Slurm on HPC). It consists of a Pilot-Manager and multiple \emph{Pilots}. The \emph{Pilot-Manager} orchestrates resources by provisioning, managing, and releasing them for workload execution. It optimizes task assignment across pilots for efficient time-to-completion and system utilization. The Pilot-Manager provides extensible resource plugins to integrate various resources.


The Pilot Abstraction's multi-level scheduling suits quantum-HPC environments for two reasons: (i) heterogeneity and complexity of quantum and classical applications complicate static workload and resource assignment, and (ii) early quantum era co-design of applications and hardware requires integrated application-level resource management. While small applications can use hard-wired integration, larger hybrid systems need more adaptive resource management. Hence, hierarchical resource management capabilities are crucial for efficiently developing and operating quantum-enabled workflow systems. Next, we present each layer of the system.




\subsubsection{L4 Workflow Layer and Abstractions:}
Pilot-Quantum provides two abstractions: \emph{(i)} resource management, enabling applications to allocate appropriate classical and quantum resources, and \emph{(ii)} workload and task management, facilitating efficient workload execution on these resources. 
For \emph{(i)}, applications must define required resources in a \texttt{pilot\_description} and submit it to Pilot-Quantum, which then acquires the resources.

For \emph{(ii)}, applications must decompose their problems into independently executable tasks, collectively forming the workload. Tasks can encapsulate units of quantum computations (e.\,g., circuit executions) and classical computations (e.\,g., pre- and post-processing). While early quantum applications require high control over problem decomposition and workload, Pilot-Quantum is expected to interface with higher-level workflow systems and application libraries. 

Tasks are defined using a \texttt{task\_description} and submitted via the \texttt{submit\_task} API. Pilot-Quantum also integrates with native abstractions of different frameworks, e.\,g.,  Dask and Ray. Tasks can be assigned (i) directly to a specific pilot or (ii) to a group of pilots. For \emph{(ii)}, the pilot manager assigns tasks to pilots, optimizing throughput by load balancing across all pilots. Please refer to \cite{pilot-quantum-api-usage} for API details.

\subsubsection{L3/L2 Workload and Task Layers:}

\IPDPS{Why is Dask used for resource management here? I understand it can be used as an executor of resources, but why not do stronger integration with, e.g., Slurm?} \alnote{second paragraph of this section}

After receiving the \texttt{pilot\_description}, the Pilot-Manager allocates requested resources via the local resource management system (\emph{steps 1-3} in Figure~\ref{fig:quantum-pilot-arch}). In HPC environments, a placeholder job, which starts the Pilot-Agent is queued via Slurm.  Once active, the Pilot Agent manages the allocated resources and executes tasks directly or through external execution engines. External execution engines are supported via plugins; currently, Pilot-Quantum supports Dask~\cite{pilot-streaming} and Ray~\cite{ray}. Further, it integrates with low-level runtimes (e.\,g., CUDA and MPI).  Quantum hardware access is handled via classical tasks that utilize, e.\,g., Qiskit Providers or PennyLane plugins.

The choice of an external execution engine depends on workload characteristics and resource infrastructure. Dask is well-suited for data-intensive workloads, such as data-parallel streaming and machine learning tasks~\cite{pilot-streaming}. Pilot-Quantum extends its capabilities with a plugin for Ray~\cite{ray}, a versatile distributed computing framework that facilitates seamless integration of GPUs (essential for quantum simulations) and custom resources like QPUs. The MPI and CUDA launch mechanisms, such as distributed state vector simulations, are crucial for tightly coupled HPC tasks. The agent leverages Slurm's \texttt{srun} command to efficiently distribute tasks across nodes, cores, and GPUs, ensuring optimal parallel execution considering machine architecture and network topology.

\subsubsection{L1 Resource Layer:}
As described, Pilot-Quantum aims to abstract heterogeneous resources through its plugin mechanism for various runtimes. In addition, it can be integrated with resource-specific tools, e.\,g., QPU-specific transpilers.

\section{Evaluation and Results}
\label{sec:eval}

\ignore{Would like to see more experiments beyond Figure 4-8; Namely, are there larger models than CIFAR-10 that are available for Qiskit and related items?}

\IPDPS{Experiment Comparisons: Comparing the results of this framework with other SOTA frameworks in the same experiments would better showcase the improvements and advantages of this approach. Some experiments appear limited in scope—e.g., only a specific quantum circuit is used in Figure 5. Providing more experimental insights and takeaways would also offer readers a clearer understanding of the results.}

This section characterizes Pilot-Quantum's capabilities using the Quantum Mini-App framework~\cite{saurabh2024quantum,QuantumMiniApp}. Mini-apps are simplified versions of quantum-HPC applications, representing key characteristics and performance bottlenecks through recurring execution patterns (motifs). We extend the circuit simulation mini-app~\cite{QuantumMiniApp} and develop a QML mini-app.

\emph{Experimental Setup:} We performed our experiments on \emph{Perlmutter}, a heterogeneous system based on the HPE Cray Shasta platform with $3,072$ CPU-only and $1,792$ GPU-accelerated nodes. GPU nodes used in our experiments comprise $64$ CPU cores and $4$ \texttt{A100} GPUs with \SI{40}{\giga\byte} resp., and \SI{80}{\giga\byte} of GPU memory. CPU nodes comprise $128$ cores with $2$\,CPUs and \SI{512}{\giga\byte} of memory. Experiments were orchestrated from a dedicated cluster node allocated via Slurm or Pilot-Quantum. This setup ensured minimal inference from other users on the Perlmutter login node.

\CCGRID{-According to Fig.2, a pilot manager must be launched on an external server outside the Perlmutter or its frontend. However, the node where the pilot manager was launched is not specified in the paper. Please specify the node where a Pilot-manager was launched in the experiments.}\alnote{addressed}




\subsection{Pilot-Quantum Characterization}
 
\begin{wraptable}[7]{R}{0.5\columnwidth}
  \centering
  \upp \upp \upp \upp
  \caption{Average Throughput and Duration by Task Size}
  \label{tbl:avg_throughput_duration}
  \upp 
  \resizebox{0.5\columnwidth}{!}
  {
  \begin{tabular}{|c|c|c|}
  \hline
  \multirowcell{1}{\emph{Tasks}} & \emph{Runtime \SI{}{[\second]}} & \emph{Throughput \SI{}{[tasks/\second]}}
  \\
  \hline
  \num{256} & \num{0.6} & \num{418.1} \\
  \hline
  \num{512}  & \num{1.3} & \num{406.5} \\ 
  \hline
  \num{1024} & \num{2.6} & \num{397.6} \\
  \hline
  \num{2048}  & \num{5.8} & \num{358.6} \\ 
  \hline
  \num{4096} & \num{11.4} & \num{358.4} \\
  \hline
  \num{8192}  & \num{27.4} & \num{298.2} \\ 
  \hline
  \end{tabular}
  }\up\up
  \end{wraptable}

Table \ref{tbl:avg_throughput_duration} shows Pilot-Quantum overhead and performance analysis for 1,024 to 8,096 zero-compute tasks across 4 Perlmutter CPU nodes (1,024 threads). The pilot job was launched using the Pilot-Quantum Slurm plugin. The task execution is handled via the Ray plugin. On average, \SI{37}{\second} were required for setting up the pilot. With increasing task counts, the Ray engine showed significant object management overhead, reducing throughput \cite{ray_overhead}. In the future, we will investigate task batching for throughput optimization.


\subsection{Circuit Simulation Mini-App}
\label{sec:circuit_simulation}

\IPDPS{There is a problem in the design of the experiments. Some experiments do not convince the advantage of the proposed method. Especially, the experiments in IV-A are simple performance evaluations of the different simulators and applications on a node. It is recommended that the performance of the simulators be compared with the performance of the simulators executed via the Pilot-Quantum.}

We extend the quantum simulation mini-app~\cite{saurabh2024quantum} by adding and evaluating different cloud backends (incl. physical quantum hardware)  (section~\ref{sec:circuit_execution}), an implementation
of distributed state vector simulation (section~\ref{sec:dist_state_vector}) and circuit cutting (section~\ref{sec:circuit_cutting}).

\subsubsection{Circuit Execution:}
\label{sec:circuit_execution}

\alnote{Mention that IONQ cloud was accessed via OLCF/ORNL}

\CCGRID{
-Experiments for Fig.3 also rack information. For example, to use IBM cloud, did you run (1) Qiskit on a node of the Perlmutter? or (2) use a system command such as  "curl"  on a node? (3) use a library to access the cloud? All these options may have different overheads, which may affect the evaluation of the efficiency of the Pilot-Quantum. Additionally, suppose the experiments are designed to evaluate the Pilot framework. In that case, these experiments should consider with and without the Pilot-Quantum,  i.e., compare simple execution times of the simulation/circuit execution and execution times the Pilot-Quantum.
}\fknote{added, please check}

\begin{figure}[t]
  \centering
  \resizebox{0.4\textwidth}{!}{
  \begin{tikzpicture}
    \begin{semilogyaxis}[
        xmin=2, xmax=30,
        ymin=0.3, ymax=40000,
        ytick={1,10,100,1000,10000},  
        xtick={4,8,12,16,20,24,28},
        grid=major,
        xlabel={Number of Qubits},
        ylabel={Time (seconds)},
        grid style={line width=.1pt, draw=gray!10},
        ymajorgrids,
        tick align=inside,
        tickwidth=0pt,
        legend columns=2,
        legend style={legend pos=north west, font=\tiny, inner xsep=4pt, outer xsep=4pt},
        legend cell align=left,
        legend image post style={xscale=0.8},
        width=0.9\linewidth, 
        height=150pt,
    ]
    
    \addplot+[
        line width=1.5pt,
        draw=s1,
        mark=square*,
        mark options={solid, fill=s1, draw=s1},
        error bars/.cd, 
            y dir=both, 
            y explicit,
        error bar style={line width=1pt, black},
        error mark options={line width=0.5pt, mark size=2pt, rotate=90, black}
    ] coordinates {
        (4, 201.5861294) +- (0, 23.57)
        (8, 237.09246) +- (0, 85.945)
        (16, 319.274479) +- (0, 36.987)
        (20, 334.282305) +- (0, 56.397)
        (22, 469.274479) +- (0, 45.34)
        (24, 475.282305) +- (0, 29.34)
        (26, 872.274479) +- (0, 72.23)
        (28, 4260) +- (0, 89.09)
    };
    \addlegendentry{IonQ}

    \addplot+[
        bar shift=5.5pt,
        dashed,
        line width=1.5pt,
        draw=s2,
        mark=*,
        mark options={solid, fill=s2, draw=s2},
        error bars/.cd, 
            y dir=both, 
            y explicit,
        error bar style={solid, line width=1pt, black},
        error mark options={line width=0.5pt, mark size=2pt, rotate=90, black}
    ] coordinates {
        (4, 241.939772) +- (0, 43.933126)
        (8, 223.079736) +- (0, 50.251054)
        (16, 245.57376) +- (0, 44.975398)
        (20, 275.459744) +- (0, 68.382344)
        (22, 198.622404) +- (0, 41.520128)
        (24, 194.127436) +- (0, 39.86133)
        (26, 197.127524) +- (0, 38.10236)
        (28, 199.123876) +- (0, 34.91847)
    };
    \addlegendentry{IBM Eagle}

    \addplot+[
        dotted,
        line width=1.5pt,
        draw=s3,
        mark=triangle*,
        mark options={solid, fill=s3, draw=s3},
        error bars/.cd, 
            y dir=both, 
            y explicit,
        error bar style={solid, line width=1pt, black},
        error mark options={line width=0.5pt, mark size=2pt, rotate=90, black}
    ] coordinates {
        (4, 39.43995333) +- (0, 4.6)
        (8, 51.32484174) +- (0, 5.3)
        (16, 63.35059524) +- (0, 6.5)
        (20, 76.18839049) +- (0, 3.9)
        (22, 88.9507823) +- (0, 4.2)
        (24, 104.6490192) +- (0, 5.1)
        (26, 168.9518147) +- (0, 7.9)
        (28, 469.7852087) +- (0, 6.8)
    };
    \addlegendentry{Aer (CPU)}
    
    \addplot+[
        dashdotted,
        line width=1.5pt,
        draw=s4,
        mark=diamond*,
        mark options={solid, fill=s4, draw=s4},
        error bars/.cd, 
            y dir=both, 
            y explicit,
        error bar style={solid, line width=1pt, black},
        error mark options={line width=0.5pt, mark size=2pt, rotate=90, black}
    ] coordinates {
        (4, 13.5861294) +- (0, 5.8)
        (8, 16.09246) +- (0, 4.5)
        (16, 21.274479) +- (0, 8.3)
        (20, 25.282305) +- (0, 10.4)
        (22, 30.274479) +- (0, 5.7)
        (24, 35.282305) +- (0, 6.5)
        (26, 42.274479) +- (0, 8.2)
        (28, 49) +- (0, 6.7)
    };
    \addlegendentry{Aer (GPU)}
    
    \end{semilogyaxis}
\end{tikzpicture}}
  \up
  \caption{\textbf{Circuit Execution on IonQ, IBM Eagle and Qiskit Aer (CPU and GPU):} Comparing the execution times for 1,024 random quantum circuits (2 to 28 qubits) on IonQ Quantum Cloud and Qiskit's Aer simulators (CPU and GPU) using Pilot-Quantum (Ray) on one Perlmutter node. We only executed eight random circuits for each qubit configuration for IBM Eagle and scaled their mean execution time to be comparable with the 1,024-circuit experiments. IonQ required 4,260 seconds for 28 qubits, while Aer (CPU) took 470 seconds, and Aer (GPU) completed it in just 49 seconds. The scaled execution time of the IBM Eagle QPU is 199 seconds.\fknote{The scaling is a bit hard to read because of the different scales on the x-axis (exp from 2 to 16 qubits and then linear for 20 to 28). Maybe we can fix the scale on the x-axis to linear and have differing separations between the data points. [I gave it a go, just comment ce\_plot\_v2 and uncomment the other figure if you want to change back] }\alnote{ok, thanks. Looks good!}} 
  \label{fig:circuit_execution_ionq_pq_scaling} 
  \upp \upp
\end{figure}
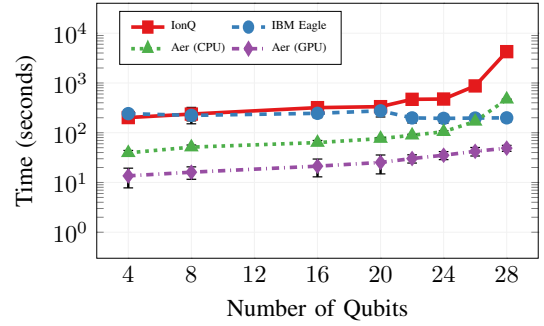


We evaluated various quantum backends, including \emph{(i)} in-process simulators using Qiskit Aer~\cite{QiskitAer}, \emph{(ii)} IonQ cloud simulators accessed via Qiskit IonQ provider~\cite{ionq_quantum_cloud}, and \emph{(iii)} IBM’s Eagle QPU via IBM cloud using Qiskit's Runtime library. We deployed Pilot-Quantum (Ray) on a Perlmutter node to manage the execution of 1,024 random circuits across these backends, ranging from $2$ to $28$ qubits.

Figure~\ref{fig:circuit_execution_ionq_pq_scaling} shows circuit execution results. For the IonQ simulator, we observe that compute time increases exponentially with qubit count, from \SI{175.8}{\second} for $2$ qubits to \SI{4260}{\second} for $28$ qubits. This aligns with the expected exponential scaling of classical quantum circuit simulation due to exponential growth in state space's size. Qiskit Aer (CPU) backend showed similar scaling but lower execution times than IonQ, with $28$ qubits taking \SI{469.7}{\second}. 
Aer (GPU) outperforms IonQ and Aer (CPU) across all qubit counts, with significantly lower execution times. For $28$ qubits, Aer (GPU) completes in \SI{49}{\second}, highlighting the advantages of GPU acceleration. Small error bars indicate stable performance, achieving an $87\times$ speedup over IonQ and a $9\times$ speedup over Aer (CPU).


To demonstrate Pilot-Quantum’s capabilities in executing quantum circuits in parallel on physical quantum hardware, we assume that 256 QPUs are available with exclusive access (i.e., no queuing time). Due to quantum hardware costs and limited availability, we execute only eight random circuits instead of 1024. The average execution time for these circuits is then scaled by a factor of 4 ($1,024/256=4$) to estimate the performance for 256 QPUs processing 1,024 random circuits. Additionally, the time accounts for the pilot quantum compute overhead of 2 seconds for 1024 circuits on a single node with 256 tasks. In contrast to the simulation experiments, the execution time varies slightly but does not increase exponentially.




\subsubsection{Distributed State Vector Simulations}\label{sec:dist_state_vector} 
enables efficient simulation of large quantum circuits with higher qubit counts and depth using multiple nodes and GPUs~\cite{DERAEDT201947, pennylane_dist_mem}. It is based  on PennyLane's \texttt{lightning.gpu}~\cite{asadi2024hybridquantumprogrammingpennylane} (\texttt{v0.41.0}), which leverage a cuQuantum~\cite{bayraktar2023cuquantum} for distributed simulation. We use $64$ nodes and $256$ GPUs.

\begin{figure}[t]
  \centering
  \resizebox{0.4\textwidth}{!}{
      \begin{tikzpicture}
    \begin{semilogyaxis}[
    tickwidth=0pt,
    grid=major,
    xlabel={Number of Qubits},
    ylabel={Time (seconds)},
    xmin=29, xmax=40,
    ymin=0.3, ymax=40000,
    xtick={29,30,...,40},
    ytick={1, 10, 100, 1000},
    grid style={line width=.1pt, draw=gray!10},
    ymajorgrids, 
    tick align=inside,
    tickwidth=0pt,
    legend columns=2,
    legend style={legend pos=north west, font=\tiny, inner xsep=4pt, outer xsep=4pt},
    legend cell align=left,
    legend image post style={xscale=0.8}, 
    width=0.9\linewidth, height=150pt,  
  ]
        \addplot[
            bar shift=5.5pt,
            dashed,
            line width=1.5pt,
            draw=s2,
            mark=*,
            mark options={solid, fill=s2, draw=s2},
            error bars/.cd, y dir=both, y explicit,
            error bar style={solid, line width=1pt, black},
            error mark options={line width=0.5pt, mark size=2pt, rotate=90, black}
        ]
        coordinates {
(30, 65.62804127)+-(0,3.03135136)
(31, 102.311)+-(0,0.2630978599)
(32, 173.6063232421875)+-(0,1.462456456)
(33, 318.5786967277527)+-(0,0.5055847774)
(34, 598.7057102733282)+-(0,0.5438289079)
        };
        \addlegendentry{(i) With Calculation}
  
        \addplot[
            bar shift=5.5pt,
            line width=1.5pt,
            draw=s1,
            mark=square*,
            mark options={solid, fill=s1, draw=s1},
            error bars/.cd, y dir=both, y explicit,
            error bar style={solid, line width=5pt, black},
            error mark options={line width=0.5pt, mark size=2pt, rotate=90, black}
        ]
        coordinates {
(30, 0.354320844) +- (0, 0.05235135523)	
(31, 0.5530284246) +- (0, 0.006685431282)	
(32, 1.048218648) +-(0, 0.02299325095)	
(33,1.976384083)+-(0,0.004751522135)	
(34,3.867900133)+- (0,0.004751522135)	
(35,7.775076787)+-(0,0.04940439002)	
(36,15.48768743)+-(0,0.01078509973)	
(37,31.03182626)+-(0,0.06436256293)	
(38,58.87557546)+-(0,0.07796759381)	
(39,117.53822)+-(0,0.01240052736)	
        };
        \addlegendentry{(ii) Without  Gradient Calculation}

        \end{semilogyaxis}
    \end{tikzpicture}}
  \up
  \caption{\textbf{PennyLane \texttt{lightning.gpu} Distributed State Vector Simulations:} Computing the expectation value of a 2-layer strongly entangling layered (SEL) circuit with and without gradient calculation using PennyLane's \texttt{lightning.gpu} device on 256 A100 40\,GB GPUs on Perlmutter. \up} 
  \label{fig:pq_mpi_dist_state_vector}
\end{figure}
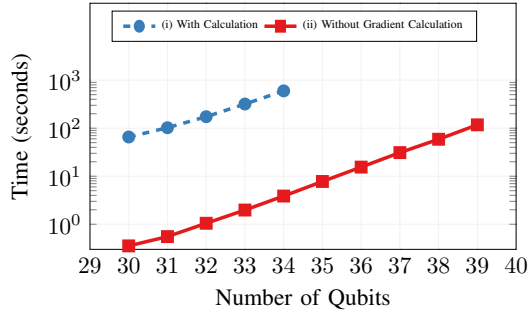

Figure~\ref{fig:pq_mpi_dist_state_vector} shows distributed state vector simulation of a 2-layer strongly entangling layered (SEL) circuit~\cite{Schuld_2020} using up to $39$ qubits. We compare circuit execution (i) with and (ii) without gradient calculation. For (i), we use adjoint differentiation~\cite{jones2020efficientcalculationgradientsclassical}, which reduces memory and compute overhead compared to finite difference methods. For (i), we could run circuits of up to $34$ qubits, and for (ii), for up to $39$ qubits.  The results underscore resource management's importance, including careful memory partitioning for state vectors and gradient data.

Figure~\ref{fig:pq_mpi_dist_state_vector} shows the distributed state vector simulation of a 2-layer strongly entangling layered (SEL) circuit~\cite{Schuld_2020}, which is frequently utilized for classification tasks. We compare circuit execution with and without gradient computation. We employ adjoint differentiation~\cite{jones2020efficientcalculationgradientsclassical} for gradient calculation, a method optimized for quantum simulations that offers lower memory and computational overhead compared to alternatives like finite difference, which necessitates multiple circuit evaluations. We investigate qubit counts from $30$ to $34$ using $256$ GPUs. The overhead of gradient calculation limits us to $34$ qubits on $256$ GPUs, while without gradient computation, we scale up to $39$ qubits. The results indicate that classical resources must be allocated appropriately between simulation and other computationally intensive tasks like ML to ensure optimal performance.


\alnote{\textbf{This is currently not observable in the figure: }
In general, we observed that every qubit doubles the GPU memory requirements. We scaled the simulations to 256 GPUs for both adjoint differentiations \& probability calculations. As we scaled the number of GPUs, the runtime of the total simulation reduced linearly.} \pnote{This is not observable, but it is a limiting factor to scale beyond 37/41 qubits as there is not enough memory https://discuss.pennylane.ai/t/lightning-gpu-failing-on-multi-node-multi-gpus/3557/14}\alnote{I meant we do not show data for runtime by number GPUs}

\alnote{future work: impact of JIT/Catalyst}
\CCGRID{Avoid using citations a subject of a sentence. E.g. [88] enables running … }
\subsubsection{Circuit Cutting~\cite{Piveteau_2024}\label{sec:circuit_cutting}}  enables running larger circuits on quantum hardware at the cost of increased sampling and classical processing overhead. It decomposes larger circuits into smaller ones, referred to as subcircuits, by cutting wires and gates, allowing execution within hardware and simulator constraints. The sampling results from the subcircuits are combined to reconstruct the original circuit outcome.  However, the number of required subcircuit executions increases exponentially with circuit size and number cuts~\cite{qiskit_circuit_cutting_sampling_overhead_table}.

This circuit-cutting motif uses Qiskit's circuit-cutting addon~\cite{qiskit-addon-cutting}. It utilizes Qiskit's \emph{EfficientSU2} ansatz, commonly used in many variational algorithms, comprising single qubit rotations and linear CNOT gates between adjacent qubits. Qiskit's expectation value reconstruction combines results from smaller sub-experiments. It provides options to run full circuit or circuit cutting simulations, enabling comparison of the circuit cutting error. It gathers comprehensive performance metrics throughout the process.


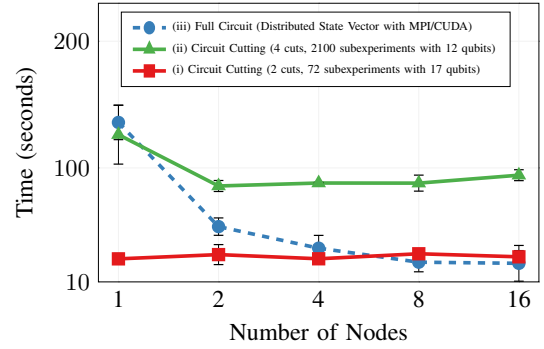
\begin{figure}[t]
  \centering
  \resizebox{0.4\textwidth}{!}{
  \begin{tikzpicture}
  \begin{axis}[
    tickwidth=0pt,
    grid=major,
    xlabel={Number of Nodes},
    ylabel={Time (seconds)},
    xmin=0.8, xmax=5.2,
    ymin=10, ymax=230,
    xtick={1, 2, 3, 4, 5},
    xticklabels={1, 2, 4, 8, 16},
    ytick={10, 100, 200},
    grid style={line width=.1pt, draw=gray!10},
    ymajorgrids,
    tick align=inside,
    tickwidth=0pt,
    legend columns=1,
    legend style={legend pos=north west, font=\tiny, inner xsep=4pt, outer xsep=4pt},
    legend cell align=left,
    legend image post style={xscale=0.8},
    width=0.9\linewidth, height=160pt
  ]

    \addplot[
      bar shift=5.5pt,
      dashed,
      line width=1.5pt,
      draw=s2,
      mark=*,
      mark options={solid, fill=s2, draw=s2},
      error bars/.cd,
        y dir=both,
        y explicit,
      error bar style={solid, line width=0.4pt, black},
      error mark options={line width=0.5pt, mark size=2pt, rotate=90, black}
    ]
    table[
      col sep=comma,
      x=ref_number,
      y=full_circuit_runtime_mean,
      y error expr=\thisrow{full_circuit_runtime_std}
    ]{experiments/circuit_cutting/34qubits_agg_fullcircuit.csv};
    \addlegendentry{(iii) Full Circuit (Distributed State Vector with MPI/CUDA)}

    \addplot[
      bar shift=5.5pt,
      line width=1.5pt,
      draw=s3,
      mark=triangle*,
      mark options={solid, fill=s3, draw=s3},
      error bars/.cd,
        y dir=both,
        y explicit,
      error bar style={solid, line width=0.4pt, black},
      error mark options={line width=0.5pt, mark size=2pt, rotate=90, black}
    ] 
    table[
      col sep=comma,
      x=ref_number,
      y=circuit_cutting_total_runtime_secs_mean_12,
      y error expr=\thisrow{circuit_cutting_total_runtime_secs_std_12}
    ]{experiments/circuit_cutting/34qubits_agg_circuit_cutting.csv};
    \addlegendentry{(ii) Circuit Cutting (4 cuts, 2100 subexperiments with 12 qubits)}

    \addplot[
      bar shift=5.5pt,
      line width=1.5pt,
      draw=s1,
      mark=square*,
      mark options={solid, fill=s1, draw=s1},
      error bars/.cd,
        y dir=both,
        y explicit,
      error bar style={solid, line width=0.4pt, black},
      error mark options={line width=0.5pt, mark size=2pt, rotate=90, black}
    ] 
    table[
      col sep=comma,
      x=ref_number,
      y=circuit_cutting_total_runtime_secs_mean_17,
      y error expr=\thisrow{circuit_cutting_total_runtime_secs_std_17}
    ]{experiments/circuit_cutting/34qubits_agg_circuit_cutting.csv};
    \addlegendentry{(i) Circuit Cutting (2 cuts, 72 subexperiments with 17 qubits)}

  \end{axis}

\end{tikzpicture}}\upp \upp
  \caption{\textbf{Qiskit Circuit Cutting technique with Pilot-Quantum:} Illustrates performance of circuit cutting for a 34 qubit circuit using 2 cuts/72 subexperiments, 4 cuts/2100 subexperiments, and full circuit simulation using Pilot-Quantum/Ray on Perlmutter 4x A100 80 GB GPUs nodes  and Qiskit's AerSimulator (with CUDA/MPI support). For circuit cutting, we use one GPU per subexperiment. For full circuit simulation, we use 4 GPUs/node and distributed state vector simulation. 
   \alnote{speedup and efficiency for parallelization of circuit cutting}\upp \upp} \upp 
  \label{fig:pq_cc_org_comparison}
\end{figure} 

We use a  $34$-qubit quantum circuit of the described ansatz in the following experiments. For circuit cutting, we utilize two configurations: \emph{(i)} subcircuit size $17$ and (ii) subcircuit size $12$. We used the \texttt{find\_cuts} method to determine the best cut and generate necessary subcircuit executions: $72$ for scenario \emph{(i)}, and approximately 2,100 subcircuit experiments for \emph{(ii)}. We compare these circuit-cutting scenarios to a full circuit simulation using Qiskit's Aer distributed state vector simulator (scenario \emph{(iii)}) with multi-node GPU and MPI support~\cite{QiskitAer,Doi_2020}. All experiments are repeated at least three times.

Figure~\ref{fig:pq_cc_org_comparison} illustrates performance scaling from $1$ to $16$ nodes ($4$ to $64$ GPUs). In scenario \emph{(i)}, with fewer subexperiments, optimal performance is achieved on a single node. Conversely, scenario \emph{(ii)} shows the best performance with $2$ nodes, achieving a~$\sim1.5\times$ speedup, demonstrating that additional resources executed the 2,100 subexperiments faster. When comparing to the sequential runtimes on one core for (i) 58 sec and (ii) 304 sec, we were able to achieve a (i) $2.1\times$  and (ii) $3.5\times$ speedup. The distributed state vector simulation (scenario (\emph{iii})) outperforms both circuit-cutting scenarios when using more than $4$ nodes. The best performance was achieved with 16 nodes, showing a $5.5\times$ speedup.



In summary, different circuit-cutting strategies lead to varying optimal performance points, highlighting the importance of application-level scheduling for efficient resource and task management, as enabled by Pilot-Quantum. Hence, an adaptive execution strategy combining multiple simulation methods (e.\,g., circuit cutting and distributed state vector simulation) and leveraging heterogeneous resources (e.\,g., CPUs, GPUs, and QPUs) will be crucial for effective application scaling.

\subsection{QML Workflow}
\label{sec:qml}

Quantum Machine Learning (QML) is a key workload~\cite{10.1145/3655027} enabling error mitigation and new applications~\cite{saurabh2023conceptual}. QML workflows typically include data preparation, training, and evaluation, leveraging parallelism for tasks like hyperparameter optimization and architecture searches. Data parallelism processes training subsets simultaneously, while tensor parallelism ensures efficient gradient computation and backpropagation. Future workflows will likely integrate QML models for inference with other tasks. This section presents a QML mini-app with a multi-stage pipeline for data encoding, compression, and classifier training using Pilot-Quantum.

\begin{figure}[t]
  \centering
    \resizebox{0.4\textwidth}{!}{
   \begin{tikzpicture}
  \begin{axis}[
    xlabel={Number of Nodes},
    grid = both,
    grid style={line width=.1pt, draw=gray!10},
    ylabel={Time (seconds)},
    axis y line*=left,
    ymin=0, ymax=25000*1.15,
    xmin=0.5, xmax=6.5,
    ymajorgrids, tick align=inside,
    tickwidth=0pt,
    xtick={1, 2, 3, 4, 5, 6},
    xticklabels={1, 2, 4, 8, 16, 32},
    ytick={0, 5000, 10000, 15000, 20000, 25000},
    legend style={at={(0.95,0.9)},anchor=north east, font=\tiny, draw=none}, %
    width=0.8\linewidth, height=150pt,
    ybar stacked, %
    scaled y ticks=false,
    bar width=15pt
  ]

    \addplot[
      color=s1,
      fill=s1,
    ] coordinates {
      (1,7314.49144499) (2,3797.77529915) (3,1930.44953205) (4,1034.75962203) (5,581.24444775) (6,475.85529953)
    };
    \addlegendentry{Sweep}
    
    \addplot[
      color=s2,
      fill=s2,
    ] coordinates {
      (1,17388.72822926) (2,9028.44484123) (3,4589.25443081) (4,2459.93231182) (5,1381.79144958) (6,1131.25000449)
    };
    \addlegendentry{BFGS}
  \end{axis}

  \begin{axis}[
    tickwidth=0pt,
    xlabel={Number of Nodes},
    ylabel={Efficiency (\%)},
    axis y line*=right,
    axis x line=none,
    ymin=0, ymax=100*1.15,
    xmin=0.5, xmax=6.5,
    xtick={1, 2, 3, 4, 5, 6},
    xticklabels={1, 2, 4, 8, 16, 32},
    ytick={0, 20, 40, 60, 80, 100},
    legend cell align={left},
    legend style={at={(0.95,0.98)},anchor=north east, font=\tiny, draw=none}, %
    width=0.8\linewidth, height=150pt
  ]

    \addplot[
      color=s3,
      line width=1.5pt, %
      error bars/.cd, 
      y dir=both, y explicit,
      error bar style={line width=1pt, black}, %
      error mark options={line width=0.5pt, mark size=2pt, rotate=90, black} %
    ] coordinates {
      (1,100) +- (0,0.25542592)
      (2,96.29968691) +- (0,0.2803805)
      (3,94.72523528) +- (0,0.95199942)
      (4,88.35979015) +- (0,3.03037082)
      (5,78.65119694) +- (0,1.89398289)
      (6,48.03516066) +- (0,2.77163905)
    };
    \addlegendentry{Efficiency}

  \end{axis}
\end{tikzpicture}}\upp \upp
  \caption{\textbf{Execution time and efficiency} for compressing the 60,000 samples of the CIFAR-10 data set for cluster configurations with varying numbers of nodes. The time is divided between the two phases, with sweeping represented by the red portion and BFGS by the blue. 
  The results demonstrate that as the number of nodes increases, the overall execution time decreases while the efficiency reduces slightly, particularly for larger node counts. 
  \up}
  \label{fig:benchmark_qml}
\end{figure}
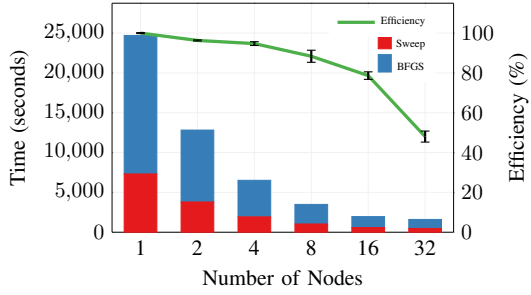

\subsubsection{Data Encoding and Compression:}
\label{sec:compression}
Efficiently representing classical data (e.g., images) for QML is critical due to quantum state preparation bottlenecks~\cite{Aaronson:2015scy, Schuld2021}. Tensor networks, like matrix product states (MPS), can efficiently approximate images~\cite{jobst2023efficientmpsrepresentationsquantum}. Our QML pipeline mini-app uses Pilot-Quantum (Ray) for a two-stage data compression process.
We evaluated the pipeline using CIFAR-10~\cite{krizhevsky2009learning} comprising $10$ classes and $60,000$ images, encoding each \(32 \times 32\) pixel image with $3$ RGB color channels into a 13-qubit quantum state ($10$ qubits encoding spatial information, and $3$ encoding color channels~\cite{rgba_encoding}). The image's quantum state is approximated by a depth $4$ circuit with 205 trainable parameters, using only single and two-qubit gates on neighboring circuits. The first stage trains the circuit using a sweeping algorithm~\cite{Rudolph_2024}, while the second stage employs parameterized circuit training with a state vector simulator and BFGS optimizer~\cite{nocedal_numerical}.


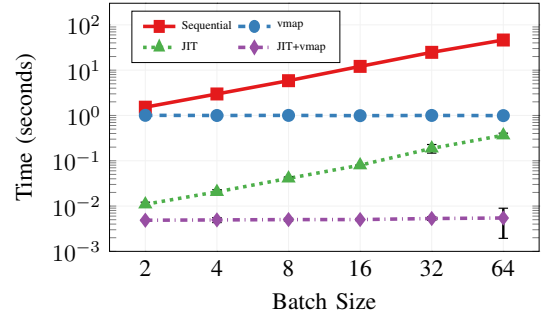
\begin{figure}[t]
  \centering
  \resizebox{0.4\textwidth}{!}{
  \begin{tikzpicture}
    \begin{semilogyaxis}[
    tickwidth=0pt,
    grid=major,
    xlabel={Batch Size},
    ylabel={Time (seconds)},
    ymin=0.001, ymax=300,
    xmin=0.5, xmax=6.5,
    xtick={1, 2, 3, 4, 5, 6},
    xticklabels={2, 4, 8, 16, 32, 64},
    ytick={1e-3, 1e-2, 1e-1, 1, 10, 100},
    grid style={line width=.1pt, draw=gray!10},
    ymajorgrids, 
    tick align=inside,
    tickwidth=0pt,
    legend columns=2,
    legend style={legend pos=north west, font=\tiny, inner xsep=4pt, outer xsep=4pt},
    legend cell align=left,
    legend image post style={xscale=0.8}, 
    width=0.9\linewidth, height=150pt
  ]
  \addplot+[
        bar shift=5.5pt,
        line width=1.5pt,
        draw=s1,
        mark=square*,
        mark options={solid, fill=s1, draw=s1},
        error bars/.cd, y dir=both, y explicit,
        error bar style={solid, line width=1pt, black},
        error mark options={line width=0.5pt, mark size=2pt, rotate=90, black}
  ] coordinates {
    (1, 1.52332445) +- (0, 0.04751922)
    (2, 2.98183821) +- (0, 0.04838781)
    (3, 5.85643724) +- (0, 0.22458307)
    (4, 12.11635395) +- (0, 0.00420862)
    (5, 24.74481913) +- (0, 0.00527187)
    (6, 46.29661712) +- (0, 0.10428157)
  };
  \addlegendentry{Sequential}
  
  \addplot+[
        bar shift=5.5pt,
        dashed,
        line width=1.5pt,
        draw=s2,
        mark=*,
        mark options={solid, fill=s2, draw=s2},
        error bars/.cd, y dir=both, y explicit,
        error bar style={solid, line width=1pt, black},
        error mark options={line width=0.5pt, mark size=2pt, rotate=90, black}
  ] coordinates {
    (1, 1.00675627) +- (0, 0.05562745)
    (2, 0.9987731) +- (0, 0.04824027)
    (3, 1.00388086) +- (0, 0.0506105)
    (4, 0.98977219) +- (0, 0.05675061)
    (5, 1.00103899) +- (0, 0.05347572)
    (6, 0.9923962) +- (0, 0.06849141)
  };
  \addlegendentry{vmap}
  
  \addplot+[
        bar shift=5.5pt,
        dotted,
        line width=1.5pt,
        draw=s3,
        mark=triangle*,
        mark options={solid, fill=s3, draw=s3},
        error bars/.cd, y dir=both, y explicit,
        error bar style={solid, line width=1pt, black},
        error mark options={line width=0.5pt, mark size=2pt, rotate=90, black}
  ] coordinates {
    (1, 0.01094554) +- (0, 0.00124416)
    (2, 0.02060177) +- (0, 0.00240222)
    (3, 0.04089269) +- (0, 0.0030415)
    (4, 0.07983058) +- (0, 0.00020308)
    (5, 0.1879533) +- (0, 0.04040736)
    (6, 0.36570323) +- (0, 0.04080451)
  };
  \addlegendentry{JIT}
  
  \addplot[
        bar shift=5.5pt,
        dashdotted,
        line width=1.5pt,
        draw=s4,
        mark=diamond*,
        mark options={solid, fill=s4, draw=s4},
        error bars/.cd, y dir=both, y explicit,
        error bar style={solid, line width=1pt, black},
        error mark options={line width=0.5pt, mark size=2pt, rotate=90, black}
  ] coordinates {
    (1, 0.00486588) +- (0, 0.000119899527)
    (2, 0.00494406) +- (0, 0.000622066564)
    (3, 0.00502804) +- (0, 0.000339465825)
    (4, 0.00503382) +- (0, 7.62132247e-05)
    (5, 0.00530925) +- (0, 0.000467020947)
    (6, 0.00545097) +- (0, 0.00351198094)
  };
  \addlegendentry{JIT+vmap}
  
  \end{semilogyaxis}
\end{tikzpicture}}\upp \upp
  \caption{\textbf{Batch processing time} by batch size for a variational quantum classifier implemented in PennyLane with JAX. The figure compares four processing methods: sequential (no optimizations), batch optimization using vmap, JIT compilation, and a combination of both. JIT compilation improves runtime by about two orders of magnitude. Batch optimization using vmap decouples the execution time from batch size. The combination of both vmap and JIT provides the best performance, significantly reducing processing time across all batch sizes. 
  }\upp \upp \upp
  \label{fig:training_loop_time}
\end{figure}




    




To demonstrate Pilot-Quantum's capabilities, we benchmarked the CIFAR-10 dataset compression runtime and efficiency across various cluster configurations (Figure~\ref{fig:benchmark_qml}). We used one ray worker node per physical node, each with $256$ logical CPUs, and observed that throughput peaked at \SI{4.5}{tasks\per\second} with $256$ CPUs. It indicates performance improvement, with up to $15\times$ speedup for the $32$-node cluster. However, efficiency decreased with increasing nodes. Efficiency, measuring task parallelization scaling with node count, is calculated as N-node cluster runtime normalized to single-node runtime and node count. The efficiency decrease stems from sequential ray worker node registration, causing initial processes to finish before the final node registers.



\subsubsection{Classification:}
\label{sec:classification}
The compressed dataset was classified using a PennyLane-implemented variational quantum classifier on one NVIDIA $A100$ GPU. The circuit uses $13$ qubits, with one arbitrary rotation layer per qubit and a ring of CNOTs~\cite{MariaSchuld2019}. Next, 10 Pauli-Z expectation values generate the prediction. The ADAM optimizer trains the model by minimizing cross-entropy loss between predicted output and target labels. We used PennyLane's \texttt{default.qubit} with JAX~\cite{jax2018github} interface, utilizing JIT and vmap for performance optimization.


Figure~\ref{fig:training_loop_time} compares batch processing times for the variational quantum classifier using four methods: sequential (unoptimized), vmap batch optimization, JIT compilation, and combined vmap-JIT. JIT compilation improves runtime by about two orders of magnitude, while vmap decouples execution time from batch size. The vmap-JIT combination performs best, significantly reducing processing time across all batch sizes. This QML mini-app demonstrates how parallelism through Pilot-Quantum (Ray) and optimization tools like JAX can greatly enhance QML workflow efficiency and scalability.

\section{Discussion and Conclusion}
\label{sec:qual_discussion}
\label{sec:discussion_conclusion}
\begin{table}[t]
  \centering
  \small
  \caption{Comparison of quantum-HPC middleware systems\upp}
  \vspace*{-0.2cm}
  \label{tbl:related}
  \resizebox{\columnwidth}{!}{
  \begin{tabular}{|@{}c@{}|@{}c@{}|@{}c@{}|@{}c@{}|@{}c@{}|}
  \hline
  \cline{2-5}
  \multirowcell{1}{\textbf{\textit{Middleware} }} & \multicolumn{1}{c|}{\textbf{Qiskit Serverless}}&\multicolumn{1}{c|}{\textbf{Covalent}}&\multicolumn{1}{c|}{\textbf{CUDA-Q}}&\multicolumn{1}{c|}{\textbf{Pilot-Quantum}}
  \\
  \cline{1-5}
  \hline\hline
  \multirowcell{1}{\textbf{\textit{Characteristics}}}& \multicolumn{4}{{c@{ }|}}{\multirowcell{1}{\textbf{\textit{L4 Workflow Layer}}}}
  \\
  \hline
  \textit{Abstraction} & Task & Lattice \& electrons & Kernel  & Task \\
   \textit{}& abstraction & abstraction (DAG)& abstraction & abstraction \\
  \hline
  \textit{Quantum} &Qiskit &\ Qiskit,\ \ &CUDA-Q &\ Qiskit,\ \ \\
  \textit{libraries} &&\ PennyLane, etc.\ \ & &\ PennyLane, etc.\ \ \\
  \cline{1-5} \hline\hline
  
  \multirowcell{1}& \multicolumn{4}{{c@{ }|}}{\multirowcell{1}{\textbf{\textit{L3 Workload Layer}}}}\\
  \hline
  \textit{Resource}  &Serverless (cloud) &Quantum, HPC,  &Local resource   & Pilot-Job \\
  \textit{management}& or container   &Cloud Executors     &management   & \\
  \textit{}          &(self-managed)     &or serverless    &(e.\,g., Slurm)  & \\
  \hline
  \textit{Task}&Application&Application&Automatic & Application\\
  \textit{partitioning}&level task&level task &partitioning (only & level task\\
  \textit{}&partitioning&partitioning &quantum tasks) &partitioning\\
  \hline
  \cline{1-5}
  \hline\hline
  \multirowcell{1}& \multicolumn{4}{{c@{ }|}}{\multirowcell{1}{\textbf{\textit{L2 Task Layer}}}}
  \\
  \hline
  \textit{Task}         &\texttt{distributed\_-}  & \texttt{electron} &None & Pilot-Job, Ray   \\
  \textit{abstraction}  &\texttt{task}  &  &  &or Dask task  \\
  \hline
  \textit{Quantum task}&same as classic &\texttt{QElectrons} &\ \texttt{cudaq.kernel} \ &same as classic\\
  \hline
  \textit{Task} &Ray &Dask &CUDA-Q  &Ray, Dask\\
  \textit{management}&&&  & \\
  \cline{1-5}
  \hline\hline
  
  \multirowcell{1}& \multicolumn{4}{{c@{ }|}}{\multirowcell{1}{\textbf{\textit{L1 Resource Layer}}}}
  \\
  \hline
  \textit{Computing } &Cloud, Docker,  &HPC, Cloud,  &HPC (Nvidia & HPC, Cloud \\
  \textit{Infrastructure}                     &Kubernetes (only  &Kubernetes  & GPUs)/ Cloud  &\\
  \textit{}& cores, no GPUs) &&(Docker-based)& \\
  \hline
  \textit{Quantum}& IBM Quantum,  & any &CUDA-Q   &any\\
  \textit{Infrastructure}        &\ Qiskit API (e.\,g.,\ \   &  & integrations &\\
  \textit{}        & IonQ, Rigetti)    &      &(e.\,g., IQM, IonQ)&\\
  \hline
  \textit{Runtime}   &Qiskit Runtime &any   &MPI, CUDA &any\\
  \hline
  \textit{Simulator} &Qiskit Aer&any & cuQuantum               &any\\
  \textit{}          &backends (e.\,g.,&    & (cuStateVec, &\\
   \textit{}          & cuQuantum) &    &  cuTensor)  &\\
  \hline
  
  
  \end{tabular}
  }
  \upp\up
  \end{table}
\subsection{Discussion}
\begin{figure}[t]
  \centering  \includegraphics[width=0.45\textwidth]{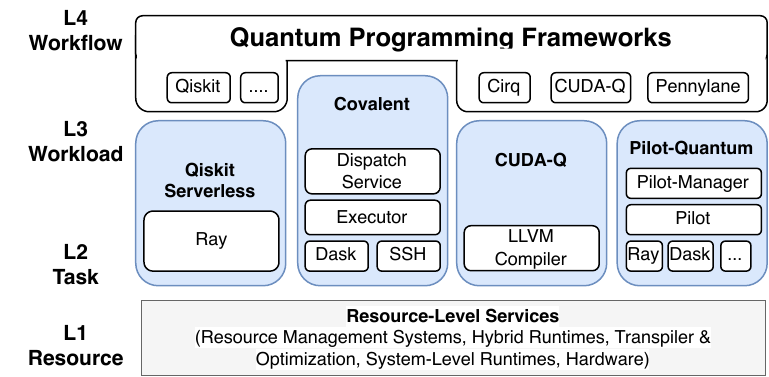}
  \caption{\textbf{Comparison of quantum-HPC Middleware: Covalent, Qiskit-Serverless, and Pilot-Quantum:} L2/L3-level middleware like Covalent, Qiskit Serverless, CUDA-Q, and Pilot-Quantum tackle deployment and scaling challenges of hybrid applications. Built on tools like CUDA, MPI, Dask, and Ray, they integrate with high-level (e.\,g., Qiskit, Cirq) and low-level software and tools (e.\,g., quantum transpilers and hybrid runtimes). \upp\upp} \upp
  \label{fig:covalent-serverless-mapping}
\end{figure}

This section qualitatively compares Pilot-Quantum with Qiskit Serverless, Covalent, and CUDA-Q by mapping each across the four architectural layers (see Figure~\ref{fig:covalent-serverless-mapping} and Table~\ref{tbl:related}). 

\emph{L4:} Qiskit Serverless and Pilot-Quantum use task-based models, while Covalent provides workflow-level abstraction at $L4$ layer. CUDA-Q's kernel-based approach supports tighter integration modes (e.\,g., error correction) and large-scale computational demands (e.\,g., distributed state vector simulations). It can be used in conjunction with Pilot-Quantum.
\IPDPS{could you explain the meaning of "new tasks can be created"? Does it mean you can spawn or launch a new task already registered? Or, you can insert a new job to your workflow?}

While each framework supports workflow dynamism, i.e., enabling new task creation based on prior results, including quantum tasks, only Covalent and Pilot-Quantum allow adding resources at runtime. At the application level, Qiskit Serverless is designed to integrate with Qiskit's libraries, while Covalent and Pilot-Quantum are framework-agnostic. In contrast, CUDA-Q is optimized for tight coupling of quantum and classical tasks. 


\emph{L3:} Both Covalent and Pilot-Quantum offer high-level control over resource allocation and task management. In contrast, Qiskit Serverless allows application-level task management but not resource allocation. At the same time, CUDA-Q uses a fixed HPC resource model with MPI and CUDA for task execution, but it lacks resource management capabilities and requires manual integration with resource management systems like Slurm.

\emph{L2:} At the task execution layer, most frameworks use task-based systems like Dask and Ray for parallel execution across diverse resources, decoupling application from execution details (e.g., scheduling). They allow some application-level control of task and resource mapping for optimally utilizing accelerator and data parallelism. Resource mapping depends heavily on application details, such as the amount of data per task, making application-level scheduling crucial for optimal performance. The CUDA-Q runtime directly manages task execution on QPUs or accelerators. While supporting some dynamism, CUDA-Q's compiled nature makes it less flexible than other task execution systems.

\emph{L1:} At the $L1$ resource layer, Qiskit Serverless, designed as a managed serverless cloud service, offers limited high-level control, e.g., the precise number of cores, memory, and GPUs cannot be specified. It integrates well with resource-specific compilers and runtimes. Covalent also provides a managed serverless service but lacks quantum-specific integrations. Both frameworks support deployment in dedicated compute environments, typically requiring manual setup with tools like Terraform. CUDA-Q does not include a cloud or cluster resource manager and assumes ownership of resources.

\emph{Summary and Limitations:} Each of the above systems targets different scenarios and quantum-HPC integration modes: Qiskit Serverless is workload-centric, Covalent focuses on workflows, CUDA-Q aims for tighter integration (\emph{HPC-for-Quantum}), while the others prioritize scaling loosely coupled workloads (\emph{Quantum-x-HPC} modes). Resource, workload, and task management are secondary for most systems. Pilot-Quantum addresses this gap by enabling seamless integration with heterogeneous resources, simplifying deployment, and supporting flexible resource use at the application level, crucial for managing diverse workloads~\cite{saurabh2024quantum}.

However, Pilot-Quantum's current implementation has limitations: it manages task dependencies at the application level, lacks high-level representations like DAGs, and restricts QPU access through classical task integration and PennyLane/Qiskit abstractions. In future work, we plan to expose QPU cloud environments through the pilot system, enabling interoperable tasks using quantum-specific languages and frameworks like OpenQASM and QIR.

\subsection{Conclusion and Future Work}
\label{sec:conclusion}

\jhanote{This last sentence can be shortened to: Pilot-Quantum will be a critical tool addressing resource and application heterogeneity as quantum computing evolves.}\nsnote{Done!: Andre: Is the below ok?} \alnote{yes}

Pilot-Quantum is middleware that enhances resource, workload, and task management in quantum-HPC systems. Pilot-Quantum is designed to handle complex hybrid applications with loosely coupled parallelism at our conceptual architecture's workload layer ($L3$). In addition, it can integrate with tightly coupled application components, e.\,g., MPI/CUDA-based distributed state vector simulations. Our experimental results demonstrate Pilot-Quantum capabilities with significant improvements in task execution times, e.g., a $3.5\times$ speedup for circuit cutting and a $15\times$ speedup for QML. Pilot-Quantum will be a critical tool for quantum-HPC workload and task management, addressing resource and application heterogeneity as quantum computing evolves.
 


In the future, we plan to support Pilot-Quantum with higher-level frameworks like Covalent and enable direct quantum resource access via OpenQASM and QIR-based tasks. We also aim to investigate scheduling algorithms for heterogeneous resource task assignments. As a future work, a key step will be developing Q-Dreamer, a predictive tool for estimating optimal resources for specific workloads and dynamically scheduling tasks based on real-time availability and system performance.





\section*{Acknowledgment}
This research used resources from the Oak Ridge Leadership Computing Facility, a DOE Office of Science User Facility supported under Contract DE-AC05-00OR22725 (Project CSC595).  Additionally, the work received support from NERSC (ERCAP0032477/ERCAP0029512: "Characterization and Middleware for Hybrid Quantum-HPC Applications") and the Bavarian State Ministry of Economic Affairs (BenchQC project, Grant DIK-0425/03). The authors used OpenAI's and Anthropic's language-generation models to generate parts of the text, which were subsequently reviewed, edited, and revised. 

\bibliographystyle{IEEEtran}
\bibliography{IEEEabrv,bibliography}{}

\begin{thebibliography}{100}
\providecommand{\url}[1]{#1}
\csname url@samestyle\endcsname
\providecommand{\newblock}{\relax}
\providecommand{\bibinfo}[2]{#2}
\providecommand{\BIBentrySTDinterwordspacing}{\spaceskip=0pt\relax}
\providecommand{\BIBentryALTinterwordstretchfactor}{4}
\providecommand{\BIBentryALTinterwordspacing}{\spaceskip=\fontdimen2\font plus
\BIBentryALTinterwordstretchfactor\fontdimen3\font minus \fontdimen4\font\relax}
\providecommand{\BIBforeignlanguage}[2]{{%
\expandafter\ifx\csname l@#1\endcsname\relax
\typeout{** WARNING: IEEEtran.bst: No hyphenation pattern has been}%
\typeout{** loaded for the language `#1'. Using the pattern for}%
\typeout{** the default language instead.}%
\else
\language=\csname l@#1\endcsname
\fi
#2}}
\providecommand{\BIBdecl}{\relax}
\BIBdecl

\bibitem{bayerstadler2021industry}
A.~Bayerstadler \emph{et~al.}, ``Industry quantum computing applications,'' \emph{EPJ Quantum Technology}, vol.~8, no.~1, p.~25, Nov 2021.

\bibitem{nielsen_chuang_2010}
M.~A. Nielsen and I.~L. Chuang, \emph{Quantum Computation and Quantum Information}.\hskip 1em plus 0.5em minus 0.4em\relax Cambridge University Press, 2010.

\bibitem{dalzell2023quantum}
A.~M. Dalzell \emph{et~al.}, ``Quantum algorithms: A survey of applications and end-to-end complexities,'' 2023.

\bibitem{quantinuum_h2}
\BIBentryALTinterwordspacing
{Quantinuum}, ``System model h2 | h-series quantum computer,'' 2024, accessed September 21, 2024. [Online]. Available: \url{https://www.quantinuum.com/products-solutions/system-model-h2-series}
\BIBentrySTDinterwordspacing

\bibitem{acharya2024quantumerrorcorrectionsurface}
\BIBentryALTinterwordspacing
R.~Acharya \emph{et~al.}, ``Quantum error correction below the surface code threshold,'' 2024. [Online]. Available: \url{https://arxiv.org/abs/2408.13687}
\BIBentrySTDinterwordspacing

\bibitem{quantumcomputingautomotiveapplications}
\BIBentryALTinterwordspacing
C.~Riofrío \emph{et~al.}, ``Quantum computing for automotive applications,'' 2024. [Online]. Available: \url{https://arxiv.org/abs/2409.14183}
\BIBentrySTDinterwordspacing

\bibitem{ALEXEEV2024666}
\BIBentryALTinterwordspacing
Y.~Alexeev \emph{et~al.}, ``Quantum-centric supercomputing for materials science: A perspective on challenges and future directions,'' \emph{Future Generation Computer Systems}, vol. 160, pp. 666--710, 2024. [Online]. Available: \url{https://www.sciencedirect.com/science/article/pii/S0167739X24002012}
\BIBentrySTDinterwordspacing

\bibitem{BECK202411}
\BIBentryALTinterwordspacing
T.~Beck \emph{et~al.}, ``Integrating quantum computing resources into scientific hpc ecosystems,'' \emph{Future Generation Computer Systems}, vol. 161, pp. 11--25, 2024. [Online]. Available: \url{https://www.sciencedirect.com/science/article/pii/S0167739X24003583}
\BIBentrySTDinterwordspacing

\bibitem{mohseni2024buildquantumsupercomputerscaling}
\BIBentryALTinterwordspacing
M.~Mohseni \emph{et~al.}, ``How to build a quantum supercomputer: Scaling challenges and opportunities,'' 2024. [Online]. Available: \url{https://arxiv.org/abs/2411.10406}
\BIBentrySTDinterwordspacing

\bibitem{RUEFENACHT2022}
M.~Ruefenacht \emph{et~al.}, ``Bringing quantum acceleration to supercomputers,'' Technical Report, \url{https://www.quantum.lrz.de/fileadmin/QIC/Downloads/IQM_HPC-QC-Integration-Whitepaper.pdf}, 2022.

\bibitem{hpcqc}
\BIBentryALTinterwordspacing
{HPC+QC Initiative}, ``Hpcqc.org -- where quantum accelerates the future of supercomputing,'' TU Munich and Danish Technical University and Leibniz Supercomputing Centre, 2024, \ Accessed on September 07, 2024. [Online]. Available: \url{https://www.hpcqc.org/}
\BIBentrySTDinterwordspacing

\bibitem{alsaadi2024exascaleworkflowapplicationsmiddleware}
\BIBentryALTinterwordspacing
A.~Alsaadi \emph{et~al.}, ``Exascale workflow applications and middleware: An exaworks retrospective,'' 2024. [Online]. Available: \url{https://arxiv.org/abs/2411.10637}
\BIBentrySTDinterwordspacing

\bibitem{elsharkawy2023integration}
A.~Elsharkawy \emph{et~al.}, ``Integration of quantum accelerators with high performance computing -- a review of quantum programming tools,'' 2023.

\bibitem{javadiabhari2024quantumcomputingqiskit}
\BIBentryALTinterwordspacing
A.~Javadi-Abhari \emph{et~al.}, ``Quantum computing with qiskit,'' 2024. [Online]. Available: \url{https://arxiv.org/abs/2405.08810}
\BIBentrySTDinterwordspacing

\bibitem{bergholm2022pennylaneautomaticdifferentiationhybrid}
\BIBentryALTinterwordspacing
V.~Bergholm \emph{et~al.}, ``Pennylane: Automatic differentiation of hybrid quantum-classical computations,'' 2022. [Online]. Available: \url{https://arxiv.org/abs/1811.04968}
\BIBentrySTDinterwordspacing

\bibitem{quantum_serverless_qsw}
\BIBentryALTinterwordspacing
I.~Faro \emph{et~al.}, ``Middleware for quantum: An orchestration of hybrid quantum-classical systems,'' in \emph{2023 IEEE International Conference on Quantum Software (QSW)}.\hskip 1em plus 0.5em minus 0.4em\relax Los Alamitos, CA, USA: IEEE Computer Society, jul 2023, pp. 1--8. [Online]. Available: \url{https://doi.ieeecomputersociety.org/10.1109/QSW59989.2023.00011}
\BIBentrySTDinterwordspacing

\bibitem{covalent_zenodo}
\BIBentryALTinterwordspacing
W.~Cunningham \emph{et~al.}, ``Agnostiqhq/covalent: v0.235.1-rc.0,'' Jun. 2024. [Online]. Available: \url{https://doi.org/10.5281/zenodo.11552147}
\BIBentrySTDinterwordspacing

\bibitem{CUDAQ}
{CUDA-Q Development Team}, ``Nvidia cuda-q for quantum-classical computing,'' \url{https://nvidia.github.io/cuda-quantum/latest/index.html}, 2024, accessed: August. 21, 2024.

\bibitem{6404423}
A.~Luckow \emph{et~al.}, ``P*: A model of pilot-abstractions,'' in \emph{2012 IEEE 8th International Conference on E-Science}, 2012, pp. 1--10.

\bibitem{10.1145/3177851}
\BIBentryALTinterwordspacing
M.~Turilli \emph{et~al.}, ``A comprehensive perspective on pilot-job systems,'' \emph{ACM Comput. Surv.}, vol.~51, no.~2, apr 2018. [Online]. Available: \url{https://doi.org/10.1145/3177851}
\BIBentrySTDinterwordspacing

\bibitem{saurabh2023conceptual}
N.~Saurabh \emph{et~al.}, ``A conceptual architecture for a quantum-hpc middleware,'' in \emph{2023 IEEE International Conference on Quantum Software (QSW)}.\hskip 1em plus 0.5em minus 0.4em\relax IEEE, 2023, pp. 116--127.

\bibitem{saurabh2024quantum}
\BIBentryALTinterwordspacing
------, ``Quantum mini-apps: A framework for developing and benchmarking quantum-hpc applications,'' in \emph{Proceedings of the 2024 Workshop on High Performance and Quantum Computing Integration}, ser. HPQCI '24.\hskip 1em plus 0.5em minus 0.4em\relax New York, NY, USA: Association for Computing Machinery, 2024, p. 11–18. [Online]. Available: \url{https://doi.org/10.1145/3659996.3660036}
\BIBentrySTDinterwordspacing

\bibitem{Qiskit_OpenSource}
{Qiskit contributors}, ``Qiskit: An open-source framework for quantum computing,'' 2023.

\bibitem{bayraktar2023cuquantum}
H.~Bayraktar \emph{et~al.}, ``cuquantum sdk: A high-performance library for accelerating quantum science,'' 2023.

\bibitem{Mantha_Pilot-Quantum_2024}
\BIBentryALTinterwordspacing
P.~Mantha \emph{et~al.}, ``{Pilot-Quantum},'' Oct. 2024. [Online]. Available: \url{https://github.com/radical-cybertools/pilot-quantum}
\BIBentrySTDinterwordspacing

\bibitem{ella2023quantumclassical}
L.~Ella \emph{et~al.}, ``Quantum-classical processing and benchmarking at the pulse-level,'' 2023.

\bibitem{hong2024entanglinglogicalqubitsbreakeven}
\BIBentryALTinterwordspacing
Y.~Hong \emph{et~al.}, ``Entangling four logical qubits beyond break-even in a nonlocal code,'' 2024. [Online]. Available: \url{https://arxiv.org/abs/2406.02666}
\BIBentrySTDinterwordspacing

\bibitem{Bluvstein_2023}
\BIBentryALTinterwordspacing
D.~Bluvstein \emph{et~al.}, ``Logical quantum processor based on reconfigurable atom arrays,'' \emph{Nature}, vol. 626, no. 7997, p. 58–65, December 2023. [Online]. Available: \url{http://dx.doi.org/10.1038/s41586-023-06927-3}
\BIBentrySTDinterwordspacing

\bibitem{PhysRevLett.127.100501}
\BIBentryALTinterwordspacing
A.~D. C\'orcoles \emph{et~al.}, ``Exploiting dynamic quantum circuits in a quantum algorithm with superconducting qubits,'' \emph{Phys. Rev. Lett.}, vol. 127, p. 100501, Aug 2021. [Online]. Available: \url{https://link.aps.org/doi/10.1103/PhysRevLett.127.100501}
\BIBentrySTDinterwordspacing

\bibitem{vqa}
\BIBentryALTinterwordspacing
M.~Cerezo \emph{et~al.}, ``Variational quantum algorithms,'' \emph{Nature Reviews Physics}, vol.~3, no.~9, pp. 625--644, 2021. [Online]. Available: \url{https://doi.org/10.1038/s42254-021-00348-9}
\BIBentrySTDinterwordspacing

\bibitem{generative_molecule_design_2022}
\BIBentryALTinterwordspacing
N.~W.~A. Gebauer \emph{et~al.}, ``Inverse design of 3d molecular structures with conditional generative neural networks,'' \emph{Nature Communications}, vol.~13, no.~1, p. 973, 2022. [Online]. Available: \url{https://doi.org/10.1038/s41467-022-28526-y}
\BIBentrySTDinterwordspacing

\bibitem{https://doi.org/10.48550/arxiv.2101.06250}
\BIBentryALTinterwordspacing
J.~Alcazar \emph{et~al.}, ``Geo: Enhancing combinatorial optimization with classical and quantum generative models,'' 2021. [Online]. Available: \url{https://arxiv.org/abs/2101.06250}
\BIBentrySTDinterwordspacing

\bibitem{Cirq2024}
\BIBentryALTinterwordspacing
{{Cirq Developers}}, ``Cirq,'' 2024, python library for quantum circuits and quantum computing. [Online]. Available: \url{https://doi.org/10.5281/zenodo.11398048}
\BIBentrySTDinterwordspacing

\bibitem{braket}
\BIBentryALTinterwordspacing
Amazon, ``{{Amazon Braket}},'' 2024. [Online]. Available: \url{https://aws.amazon.com/braket/}
\BIBentrySTDinterwordspacing

\bibitem{seidel2024qrispframeworkcompilablehighlevel}
\BIBentryALTinterwordspacing
R.~Seidel \emph{et~al.}, ``Qrisp: A framework for compilable high-level programming of gate-based quantum computers,'' 2024. [Online]. Available: \url{https://arxiv.org/abs/2406.14792}
\BIBentrySTDinterwordspacing

\bibitem{10.1145/3385412.3386007}
\BIBentryALTinterwordspacing
B.~Bichsel \emph{et~al.}, ``Silq: a high-level quantum language with safe uncomputation and intuitive semantics,'' in \emph{Proceedings of the 41st ACM SIGPLAN Conference on Programming Language Design and Implementation}, ser. PLDI 2020.\hskip 1em plus 0.5em minus 0.4em\relax New York, NY, USA: Association for Computing Machinery, 2020, p. 286–300. [Online]. Available: \url{https://doi.org/10.1145/3385412.3386007}
\BIBentrySTDinterwordspacing

\bibitem{Svore_2018}
\BIBentryALTinterwordspacing
K.~Svore \emph{et~al.}, ``Q\#: Enabling scalable quantum computing and development with a high-level dsl,'' in \emph{Proceedings of the Real World Domain Specific Languages Workshop 2018}, ser. RWDSL2018.\hskip 1em plus 0.5em minus 0.4em\relax ACM, Feb. 2018. [Online]. Available: \url{http://dx.doi.org/10.1145/3183895.3183901}
\BIBentrySTDinterwordspacing

\bibitem{qiskit_optimization}
{Qiskit Community}, ``Qiskit optimization,'' \url{https://qiskit-community.github.io/qiskit-optimization/}, 2024, accessed: 01 October 2024.

\bibitem{qiskit_nature}
------, ``Qiskit nature,'' \url{https://qiskit-community.github.io/qiskit-nature/}, 2024, accessed: 01 October 2024.

\bibitem{mcclean2019openfermionelectronicstructurepackage}
\BIBentryALTinterwordspacing
J.~R. McClean \emph{et~al.}, ``Openfermion: The electronic structure package for quantum computers,'' 2019. [Online]. Available: \url{https://arxiv.org/abs/1710.07629}
\BIBentrySTDinterwordspacing

\bibitem{Qiskit_IBM_Runtime}
``Ibm client for qiskit runtime,'' \url{https://github.com/Qiskit/qiskit-ibm-runtime}, 2023, accessed: 2023-12-31.

\bibitem{AWSBraketJobs}
\BIBentryALTinterwordspacing
{Amazon Web Services}. (2024) Managing your amazon braket hybrid job. Amazon Web Services. [Online]. Available: \url{https://docs.aws.amazon.com/braket/latest/developerguide/braket-jobs.html}
\BIBentrySTDinterwordspacing

\bibitem{braket-jobs-2021}
D.~Poccia, ``Introducing amazon braket hybrid jobs – set up, monitor, and efficiently run hybrid quantum-classical workloads,'' \url{https://aws.amazon.com/blogs/aws/introducing-amazon-braket-hybrid-jobs-set-up-monitor-and-efficiently-run-hybrid-quantum-classical-workloads/}, 2021.

\bibitem{nakaji2024generative}
K.~Nakaji \emph{et~al.}, ``The generative quantum eigensolver (gqe) and its application for ground state search,'' 2024.

\bibitem{xacc_2020}
\BIBentryALTinterwordspacing
A.~J. McCaskey \emph{et~al.}, ``{XACC}: a system-level software infrastructure for heterogeneous quantum{\textendash}classical computing,'' \emph{Quantum Science and Technology}, vol.~5, no.~2, p. 024002, feb 2020. [Online]. Available: \url{https://doi.org/10.1088%2F2058-9565%2Fab6bf6}
\BIBentrySTDinterwordspacing

\bibitem{nguyen2020extendingcheterogeneousquantumclassical}
\BIBentryALTinterwordspacing
T.~Nguyen \emph{et~al.}, ``Extending c++ for heterogeneous quantum-classical computing,'' 2020. [Online]. Available: \url{https://arxiv.org/abs/2010.03935}
\BIBentrySTDinterwordspacing

\bibitem{QiskitBackends2023}
\BIBentryALTinterwordspacing
E.~Winston and D.~Moreda, ``Qiskit backends: What they are and how to work with them,'' \emph{Medium}, 2018. [Online]. Available: \url{https://medium.com/qiskit/qiskit-backends-what-they-are-and-how-to-work-with-them-fb66b3bd0463}
\BIBentrySTDinterwordspacing

\bibitem{pennylane_plugins}
\BIBentryALTinterwordspacing
{Xanadu}, ``Pennylane plugins,'' 2024, accessed on October 01, 2024. [Online]. Available: \url{https://pennylane.ai/plugins/}
\BIBentrySTDinterwordspacing

\bibitem{Cross_2022}
\BIBentryALTinterwordspacing
A.~Cross \emph{et~al.}, ``Openqasm 3: A broader and deeper quantum assembly language,'' \emph{ACM Transactions on Quantum Computing}, vol.~3, no.~3, p. 1–50, Sep. 2022. [Online]. Available: \url{http://dx.doi.org/10.1145/3505636}
\BIBentrySTDinterwordspacing

\bibitem{QIRAlliance}
\BIBentryALTinterwordspacing
{QIR Alliance}, ``Qir alliance: The core of quantum development,'' \url{https://www.qir-alliance.org/}, 2024, accessed on September 29, 2024. [Online]. Available: \url{https://www.qir-alliance.org/}
\BIBentrySTDinterwordspacing

\bibitem{lubinski2022advancinghybridquantumclassicalcomputation}
\BIBentryALTinterwordspacing
T.~Lubinski \emph{et~al.}, ``Advancing hybrid quantum-classical computation with real-time execution,'' 2022. [Online]. Available: \url{https://arxiv.org/abs/2206.12950}
\BIBentrySTDinterwordspacing

\bibitem{tket}
\BIBentryALTinterwordspacing
S.~Sivarajah \emph{et~al.}, ``tket: a retargetable compiler for nisq devices,'' \emph{Quantum Science and Technology}, vol.~6, no.~1, p. 014003, nov 2020. [Online]. Available: \url{https://dx.doi.org/10.1088/2058-9565/ab8e92}
\BIBentrySTDinterwordspacing

\bibitem{osti_1785933}
\BIBentryALTinterwordspacing
E.~Younis \emph{et~al.}, ``Berkeley quantum synthesis toolkit (bqskit) v1,'' 4 2021. [Online]. Available: \url{https://www.osti.gov//servlets/purl/1785933}
\BIBentrySTDinterwordspacing

\bibitem{staq}
\BIBentryALTinterwordspacing
M.~Amy and V.~Gheorghiu, ``staq—a full-stack quantum processing toolkit,'' \emph{Quantum Science and Technology}, vol.~5, no.~3, p. 034016, jun 2020. [Online]. Available: \url{https://dx.doi.org/10.1088/2058-9565/ab9359}
\BIBentrySTDinterwordspacing

\bibitem{elsharkawy2024integrationquantumacceleratorshpc}
\BIBentryALTinterwordspacing
A.~Elsharkawy \emph{et~al.}, ``Integration of quantum accelerators into hpc: Toward a unified quantum platform,'' 2024. [Online]. Available: \url{https://arxiv.org/abs/2407.18527}
\BIBentrySTDinterwordspacing

\bibitem{AHN2020202}
\BIBentryALTinterwordspacing
D.~H. Ahn \emph{et~al.}, ``Flux: Overcoming scheduling challenges for exascale workflows,'' \emph{Future Generation Computer Systems}, vol. 110, pp. 202--213, 2020. [Online]. Available: \url{https://www.sciencedirect.com/science/article/pii/S0167739X19317169}
\BIBentrySTDinterwordspacing

\bibitem{288717}
\BIBentryALTinterwordspacing
Q.~Weng \emph{et~al.}, ``Beware of fragmentation: Scheduling {GPU-Sharing} workloads with fragmentation gradient descent,'' in \emph{2023 USENIX Annual Technical Conference (USENIX ATC 23)}.\hskip 1em plus 0.5em minus 0.4em\relax Boston, MA: USENIX Association, Jul. 2023, pp. 995--1008. [Online]. Available: \url{https://www.usenix.org/conference/atc23/presentation/weng}
\BIBentrySTDinterwordspacing

\bibitem{nguyen2024qfaas}
H.~T. Nguyen \emph{et~al.}, ``Qfaas: A serverless function-as-a-service framework for quantum computing,'' \emph{Future Generation Computer Systems}, vol. 154, pp. 281--300, 2024.

\bibitem{https://doi.org/10.48550/arxiv.2211.02350}
\BIBentryALTinterwordspacing
S.~Sivarajah \emph{et~al.}, ``Tierkreis: A dataflow framework for hybrid quantum-classical computing,'' 2022. [Online]. Available: \url{https://arxiv.org/abs/2211.02350}
\BIBentrySTDinterwordspacing

\bibitem{qibo_paper}
\BIBentryALTinterwordspacing
S.~Efthymiou \emph{et~al.}, ``Qibo: a framework for quantum simulation with hardware acceleration,'' \emph{Quantum Science and Technology}, vol.~7, no.~1, p. 015018, dec 2021. [Online]. Available: \url{https://doi.org/10.1088/2058-9565/ac39f5}
\BIBentrySTDinterwordspacing

\bibitem{shehata2024frameworkintegratingquantumsimulation}
\BIBentryALTinterwordspacing
A.~Shehata \emph{et~al.}, ``A framework for integrating quantum simulation and high performance computing,'' 2024. [Online]. Available: \url{https://arxiv.org/abs/2408.08098}
\BIBentrySTDinterwordspacing

\bibitem{Nguyen2024}
\BIBentryALTinterwordspacing
T.~Nguyen \emph{et~al.}, \emph{Software for Massively Parallel Quantum Computing}.\hskip 1em plus 0.5em minus 0.4em\relax Cham: Springer International Publishing, 2024, pp. 101--119. [Online]. Available: \url{https://doi.org/10.1007/978-3-031-37966-6_6}
\BIBentrySTDinterwordspacing

\bibitem{rasqal}
{Rasqal}, ``Quantum-classical runtime solver,'' \url{https://github.com/oqc-community/rasqal}, 2024, \ Accessed: 2024-12-19.

\bibitem{zapata2021orchestra}
{{Zapata Computing}}, ``Orquestra: A platform for hybrid quantum-classical computing,'' Zapata Computing, \url{https://www.zapatacomputing.com/orquestra-platform/}, 2023.

\bibitem{ibm_quantum_serverless_2023}
{IBM Quantum}, ``Qiskit patterns with quantum serverless,'' \url{https://docs.quantum.ibm.com/run/quantum-serverless#qiskit-patterns-with-quantum-serverless}, 2023, accessed: 2024-10-04.

\bibitem{IBMQiskitServerless}
\BIBentryALTinterwordspacing
------. (2024) Qiskit serverless sets the stage for qiskit functions. IBM. [Online]. Available: \url{https://www.ibm.com/quantum/blog/qiskit-serverless}
\BIBentrySTDinterwordspacing

\bibitem{quantum_serverless_0.17.0}
``Qiskit serverless 0.18.1,'' \url{https://github.com/Qiskit/qiskit-serverless}, accessed: 2024-12-23.

\bibitem{ray}
\BIBentryALTinterwordspacing
P.~Moritz \emph{et~al.}, ``Ray: A distributed framework for emerging {AI} applications,'' in \emph{13th USENIX Symposium on Operating Systems Design and Implementation (OSDI 18)}.\hskip 1em plus 0.5em minus 0.4em\relax Carlsbad, CA: USENIX Association, October 2018, pp. 561--577. [Online]. Available: \url{https://www.usenix.org/conference/osdi18/presentation/moritz}
\BIBentrySTDinterwordspacing

\bibitem{qiskit_ionq_provider}
\BIBentryALTinterwordspacing
{{Qiskit Community}}, ``Qiskit ionq provider,'' 2024, accessed: 2024-10-04. [Online]. Available: \url{https://github.com/qiskit-community/qiskit-ionq}
\BIBentrySTDinterwordspacing

\bibitem{qiskit_braket_provider}
\BIBentryALTinterwordspacing
------, ``Qiskit braket provider,'' 2024, accessed: 2024-10-04. [Online]. Available: \url{https://github.com/qiskit-community/qiskit-braket-provider}
\BIBentrySTDinterwordspacing

\bibitem{covalent2023}
{{Covalent}}, ``Covalent: Open source workflow orchestration for heterogenous computing,'' Covalent, \url{https://www.covalent.xyz/}, 2023.

\bibitem{dask}
{Dask Development Team}, ``Dask: Scale the python tools you love,'' \url{https://www.dask.org}, 2024, accessed: Apr. 21, 2024.

\bibitem{covalent_hpc_plugin}
{AgnostiqHQ}, ``Covalent hpc plugin: Covalent plugin for hpc batch job schedulers,'' \url{https://github.com/Quantum-Accelerators/covalent-hpc-plugin}, 2023.

\bibitem{llvm}
\BIBentryALTinterwordspacing
{LLVM Project}, ``{The LLVM Compiler Infrastructure Project},'' 2024, accessed: 2024-10-05. [Online]. Available: \url{https://llvm.org/}
\BIBentrySTDinterwordspacing

\bibitem{10.1145/369028.369109}
\BIBentryALTinterwordspacing
F.~D. Berman \emph{et~al.}, ``Application-level scheduling on distributed heterogeneous networks,'' in \emph{Proceedings of the 1996 ACM/IEEE Conference on Supercomputing}, ser. Supercomputing '96.\hskip 1em plus 0.5em minus 0.4em\relax USA: IEEE Computer Society, 1996, p. 39–es. [Online]. Available: \url{https://doi.org/10.1145/369028.369109}
\BIBentrySTDinterwordspacing

\bibitem{QuantumMiniApp}
\BIBentryALTinterwordspacing
N.~Saurabh \emph{et~al.}, ``{Quantum Mini-Apps},'' Jun. 2024. [Online]. Available: \url{https://github.com/radical-cybertools/quantum-mini-apps}
\BIBentrySTDinterwordspacing

\bibitem{9309042}
T.~Ben-Nun \emph{et~al.}, ``Workflows are the new applications: Challenges in performance, portability, and productivity,'' in \emph{2020 IEEE/ACM International Workshop on Performance, Portability and Productivity in HPC (P3HPC)}, 2020, pp. 57--69.

\bibitem{pilot-quantum-api-usage}
{Pradeep Mantha, Andre Luckow}, ``Pilot quantum api usage,'' \url{https://github.com/radical-cybertools/pilot-quantum?tab=readme-ov-file#api-usage}, 2024, \ Accessed: 2024-12-22.

\bibitem{pilot-streaming}
G.~Chantzialexiou \emph{et~al.}, ``Pilot-streaming: A stream processing framework for high-performance computing,'' in \emph{2018 IEEE 14th International Conference on e-Science (e-Science)}, 2018, pp. 177--188.

\bibitem{ray_overhead}
{Ray Team}, ``Anti-pattern: Over-parallelizing with too fine-grained tasks harms speedup,'' \url{https://docs.ray.io/en/latest/ray-core/patterns/too-fine-grained-tasks.html}, 2024, \ Accessed: 2024-12-22.

\bibitem{QiskitAer}
\BIBentryALTinterwordspacing
{Qiskit Team}, ``Qiskit aer documentation,'' IBM, 2024, high-performance quantum computing simulators with realistic noise models. [Online]. Available: \url{https://qiskit.github.io/qiskit-aer/}
\BIBentrySTDinterwordspacing

\bibitem{ionq_quantum_cloud}
\BIBentryALTinterwordspacing
{IonQ}, ``Ionq quantum cloud,'' \url{https://ionq.com/quantum-cloud}, 2024, accessed on October 06, 2024. [Online]. Available: \url{https://ionq.com/quantum-cloud}
\BIBentrySTDinterwordspacing

\bibitem{DERAEDT201947}
\BIBentryALTinterwordspacing
H.~{De Raedt} \emph{et~al.}, ``Massively parallel quantum computer simulator, eleven years later,'' \emph{Computer Physics Communications}, vol. 237, pp. 47--61, 2019. [Online]. Available: \url{https://www.sciencedirect.com/science/article/pii/S0010465518303977}
\BIBentrySTDinterwordspacing

\bibitem{pennylane_dist_mem}
``Distributing quantum simulations using lightning.gpu with nvidia cuquantum,'' \url{https://pennylane.ai/blog/2023/09/distributing-quantum-simulations-using-lightning-gpu-with-NVIDIA-cuQuantum/}, accessed: 2023-09-12.

\bibitem{asadi2024hybridquantumprogrammingpennylane}
\BIBentryALTinterwordspacing
A.~Asadi \emph{et~al.}, ``Hybrid quantum programming with pennylane lightning on hpc platforms,'' 2024. [Online]. Available: \url{https://arxiv.org/abs/2403.02512}
\BIBentrySTDinterwordspacing

\bibitem{Schuld_2020}
\BIBentryALTinterwordspacing
M.~Schuld \emph{et~al.}, ``Circuit-centric quantum classifiers,'' \emph{Physical Review A}, vol. 101, no.~3, Mar. 2020. [Online]. Available: \url{http://dx.doi.org/10.1103/PhysRevA.101.032308}
\BIBentrySTDinterwordspacing

\bibitem{jones2020efficientcalculationgradientsclassical}
\BIBentryALTinterwordspacing
T.~Jones and J.~Gacon, ``Efficient calculation of gradients in classical simulations of variational quantum algorithms,'' 2020. [Online]. Available: \url{https://arxiv.org/abs/2009.02823}
\BIBentrySTDinterwordspacing

\bibitem{Piveteau_2024}
\BIBentryALTinterwordspacing
C.~Piveteau and D.~Sutter, ``Circuit knitting with classical communication,'' \emph{IEEE Transactions on Information Theory}, vol.~70, no.~4, p. 2734–2745, Apr. 2024. [Online]. Available: \url{http://dx.doi.org/10.1109/TIT.2023.3310797}
\BIBentrySTDinterwordspacing

\bibitem{qiskit_circuit_cutting_sampling_overhead_table}
\BIBentryALTinterwordspacing
{IBM}, ``Sampling overhead reference table,'' 2024, accessed on October 01, 2024. [Online]. Available: \url{https://qiskit.github.io/qiskit-addon-cutting/explanation/index.html#sampling-overhead-reference-table}
\BIBentrySTDinterwordspacing

\bibitem{qiskit-addon-cutting}
A.~M. Bra\'{n}czyk \emph{et~al.}, ``{Qiskit addon: circuit cutting},'' \url{https://github.com/Qiskit/qiskit-addon-cutting}, 2024.

\bibitem{Doi_2020}
\BIBentryALTinterwordspacing
J.~Doi and H.~Horii, ``Cache blocking technique to large scale quantum computing simulation on supercomputers,'' in \emph{2020 IEEE International Conference on Quantum Computing and Engineering (QCE)}.\hskip 1em plus 0.5em minus 0.4em\relax IEEE, Oct. 2020. [Online]. Available: \url{http://dx.doi.org/10.1109/QCE49297.2020.00035}
\BIBentrySTDinterwordspacing

\bibitem{10.1145/3655027}
\BIBentryALTinterwordspacing
C.~A. Riofrio \emph{et~al.}, ``A characterization of quantum generative models,'' \emph{ACM Transactions on Quantum Computing}, vol.~5, no.~2, jun 2024. [Online]. Available: \url{https://doi.org/10.1145/3655027}
\BIBentrySTDinterwordspacing

\bibitem{Aaronson:2015scy}
S.~Aaronson, ``{Read the fine print},'' \emph{Nature Phys.}, vol.~11, no.~4, pp. 291--293, 2015.

\bibitem{Schuld2021}
\BIBentryALTinterwordspacing
M.~Schuld and F.~Petruccione, \emph{Machine Learning with Quantum Computers}, ser. Quantum Science and Technology.\hskip 1em plus 0.5em minus 0.4em\relax Springer International Publishing, 2021. [Online]. Available: \url{https://books.google.de/books?id=-N5IEAAAQBAJ}
\BIBentrySTDinterwordspacing

\bibitem{jobst2023efficientmpsrepresentationsquantum}
\BIBentryALTinterwordspacing
B.~Jobst \emph{et~al.}, ``Efficient mps representations and quantum circuits from the fourier modes of classical image data,'' 2023. [Online]. Available: \url{https://arxiv.org/abs/2311.07666}
\BIBentrySTDinterwordspacing

\bibitem{krizhevsky2009learning}
\BIBentryALTinterwordspacing
A.~Krizhevsky, ``Learning multiple layers of features from tiny images,'' pp. 32--33, 2009. [Online]. Available: \url{https://www.cs.toronto.edu/~kriz/learning-features-2009-TR.pdf}
\BIBentrySTDinterwordspacing

\bibitem{rgba_encoding}
B.~Sun \emph{et~al.}, ``A multichannel representation for images on quantum computers using the rgb$\alpha$ color space,'' \emph{WISP 2011 - IEEE International Symposium on Intelligent Signal Processing}, 09 2011.

\bibitem{Rudolph_2024}
\BIBentryALTinterwordspacing
M.~S. Rudolph \emph{et~al.}, ``Decomposition of matrix product states into shallow quantum circuits,'' \emph{Quantum Science and Technology}, vol.~9, no.~1, p. 015012, nov 2023. [Online]. Available: \url{https://dx.doi.org/10.1088/2058-9565/ad04e6}
\BIBentrySTDinterwordspacing

\bibitem{nocedal_numerical}
J.~Nocedal and S.~Wright, \emph{Numerical optimization}, 2nd~ed., ser. Springer series in operations research and financial engineering.\hskip 1em plus 0.5em minus 0.4em\relax New York, NY: Springer, 2006.

\bibitem{MariaSchuld2019}
M.~Schuld, ``Variational classifier,'' \url{https://pennylane.ai/qml/demos/tutorial_variational_classifier/}, 10 2019, date Accessed: 2024-10-09.

\bibitem{jax2018github}
\BIBentryALTinterwordspacing
J.~Bradbury \emph{et~al.}, ``{JAX}: composable transformations of {P}ython+{N}um{P}y programs,'' 2018. [Online]. Available: \url{http://github.com/jax-ml/jax}
\BIBentrySTDinterwordspacing

\end{thebibliography}

\end{document}